\documentclass[12pt]{article}

\AtBeginDocument{
  \addtocontents{toc}{\normalsize}
}

\pdfoutput=1

\usepackage{amsmath,amssymb,amsfonts,amsthm,amscd,mathrsfs}
\usepackage{xcolor}
\definecolor{darkblue}{rgb}{0.1,0.1,.7}
\usepackage[colorlinks, linkcolor=darkblue, citecolor=darkblue, urlcolor=darkblue, linktocpage]{hyperref} 
\usepackage[square, comma, compress,numbers]{natbib}
\usepackage[]{version}
\usepackage[]{graphicx}
\usepackage[]{latexsym}
\usepackage{geometry}
\usepackage[all,cmtip]{xy}
\usepackage[margin=10pt,font=small,labelfont=bf]{caption}
\usepackage{ifthen}
\usepackage{tikz}
\usepackage{array,setspace,mathrsfs,amsfonts,yfonts,dsfont,bbm,colonequals}
\usepackage{dsfont} 
\usepackage{xspace}

\usepackage{accents}
\newlength{\dhatheight}

\newcommand{\reef}[1]{(\ref{#1})}
\def\vareps{\varepsilon}
\def\eps{\epsilon}
\newcommand{\beq}{\begin{equation}} 
\newcommand{\eeq}{\end{equation}}
\def\del {\partial} 
\def\nn{\nonumber} 
\def\bZ {\mathbb{Z}} 
\def\bR {\mathbb{R}} 
 
\def\calO {{\cal O}} 
 
\def\calD {{\cal D}} 
 
\def\calN {{\cal N}} 
\def\calM {{\cal M}}

\def\calC {{\cal C}} 
\def\bZ {\mathbb{Z}} 
 
\def\half{{\textstyle\frac 12}}

\def\ge{\geqslant}
\def\le{\leqslant}

\def\nn{\nonumber}
\def \del{\partial}
\def\Sd{\mathrm{S}_d}
\def\eps{\epsilon}
\def\unit{\mathds{1}} 
\newcommand{\NO}[1]{{:\!#1\!:}}
\renewcommand{\d}{{\rm d}}

\newcommand{\fl}{\mathcal{L}_\sigma}

\newcommand{\op}{\ensuremath{\mathcal{O}}\xspace}

\newcommand{\vev}[1]{\ensuremath{\langle #1 \rangle}\xspace}

\newcommand{\hf}{\frac{1}{2}}

\newcommand{\G}{\Gamma}

\let\D=\Delta
\let\e=\epsilon

\newcommand{\be}{\begin{equation}}
\newcommand{\ee}{\end{equation}}
\def\ba{\begin{array}}
\def\ea{\end{array}}

\newcommand{\bea}{\begin{eqnarray}}
\newcommand{\eea}{\end{eqnarray}}

\newcommand{\ud}{\mathrm d}

\numberwithin{equation}{section}
\begin{document}

\vspace*{-.6in} \thispagestyle{empty}
\begin{flushright}
CERN PH-TH/2015-200\\
\end{flushright}
\vspace{1cm} {\Large
\begin{center}
{\bf Conformal Invariance in the Long-Range Ising Model}\\
\end{center}}
\vspace{1cm}
\begin{center}
{\bf Miguel F.~Paulos$^{a}$, Slava Rychkov$^{a,b,c}$, Balt C. van Rees$^a$, Bernardo Zan$^{d}$}\\[2cm] 
{
\small
$^{a}$ CERN, Theory Group, Geneva, Switzerland\\
$^{b}$ Laboratoire de Physique Th\'{e}orique de l'\'{E}cole Normale Sup\'{e}rieure (LPTENS), Paris, France
\\
$^{c}$ Facult\'e de Physique, Universit\'{e} Pierre et Marie Curie (UPMC), Paris, France 
\\
$^{d}$ Institute of Physics, Universiteit van Amsterdam, Amsterdam, The Netherlands
\normalsize
}
\\
\end{center}

\vspace{4mm}

\begin{abstract}
We consider the question of conformal invariance of the long-range Ising model at the critical point. The continuum description is given in terms of a nonlocal field theory, and the absence of a stress tensor invalidates all of the standard arguments for the enhancement of scale invariance to conformal invariance. We however show that several correlation functions, computed to second order in the epsilon expansion, are nontrivially consistent with conformal invariance. We proceed to give a proof of conformal invariance to all orders in the epsilon expansion, based on the description of the long-range Ising model as a defect theory in an auxiliary higher-dimensional space. A detailed review of conformal invariance in the $d$-dimensional short-range Ising model is also included and may be of independent interest.

\end{abstract}
\vspace{.2in}
\vspace{.3in}

\hspace{0.7cm} August 2015

\newpage

{
\setlength{\parskip}{0.05in}
\renewcommand{\baselinestretch}{0.7}\normalsize
\tableofcontents
\renewcommand{\baselinestretch}{1.0}\normalsize
}


\setlength{\parskip}{0.1in}
\newpage

\section{Introduction} 

We will be studying the long-range Ising (LRI) model---a close cousin of the usual, short-range, Ising model (SRI). While in the SRI only nearest-neighbor spins interact, the LRI energy involves interactions at arbitrary distances with a powerlaw potential:
\beq
H_{\rm LRI}=-J\sum_{i,j} s_i s_j/ r_{ij}^{d+\sigma}\,,
\eeq
where $d$ is the space dimensionality, and $\sigma>0$ is a free parameter. The sum is over all pairs of lattice sites and $r_{ij}$ is the distance between them. We are considering the ferromagnetic interaction case $J>0$.

\begin{figure}[htbp]
\begin{center}
\includegraphics[scale=0.8]{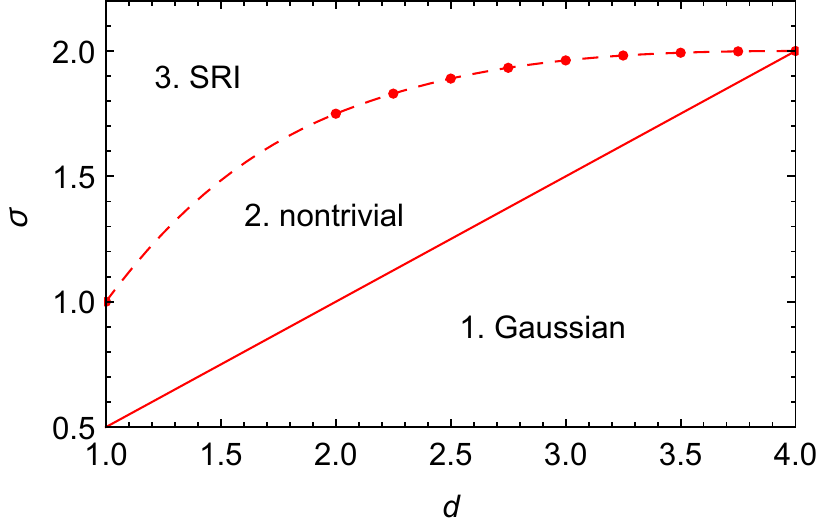}
\caption{The three phases for the LRI critical point. The boundaries are $\sigma=d/2$ (straight solid) and $\sigma=2-\eta_{\rm SRI}(d)$ (curved dashed, interpolated using the known exact $\eta_{\rm SRI}(d)$ for $d=1,2,4$ and numerical values from the $\eps$-expansion \cite{LeGuillou:1987ph} and the conformal bootstrap \cite{El-Showk:2013nia} for a few intermediate $d$). Here we will be working near the boundary $\sigma=d/2$.}
\label{fig-standard}
\end{center}
\end{figure}

Just like the SRI, the LRI has a second-order phase transition at a critical temperature $T=T_c$. The critical theory is universal, i.e.~independent of the short-distance details such as the choice of the lattice. It has however an interesting dependence on $\sigma$ (see Fig.~\ref{fig-standard}):

1. For $\sigma<d/2$, the critical point is a gaussian theory described by the nonlocal action involving the ``fractional Laplacian'' operator $\fl \equiv (-\,\del^2)^{\sigma/2}$:
\beq
S_0=\int {\rm d}^dx\, \d^d y\, \phi(x)\phi(y)/|x-y|^{d+\sigma}\,\propto \int {\rm d}^dx\, \phi\, \fl \phi \,.
\label{eq:kernel}
\eeq
The field $\phi$ represents the spin density, and its scaling dimension is read off as
\beq
\Delta_\phi=(d-\sigma)/2\,.
\label{eq:[phi]}
\eeq
Composite operators can be built out of $\phi$ by differentiations and taking normal-ordered products. Since the theory is gaussian, there are no anomalous dimensions.
 
2. For $d/2<\sigma<\sigma_*$, the critical point is a nontrivial, non-gaussian, theory, whose field-theoretic description can be obtained by perturbing the nonlocal action $S_0$ by a local quartic interaction $\int {\rm d}^dx\, \phi^4(x)$. In the range $\sigma>d/2$ this interaction is relevant and generates a renormalization group (RG) flow, reaching a fixed point in the IR. This IR fixed point is believed to be in the same universality class as the critical point of the LRI lattice model. Interestingly, as it will be explained below, the dimension of $\phi$ at the fixed point is still given by the same formula \reef{eq:[phi]}. However, composite operators do get nontrivial anomalous dimensions.

3. Finally, for $\sigma>\sigma_*$ the potential is so strongly peaked at short distances that the LRI critical point is identical with the SRI critical point, i.e.~is independent of $\sigma$. The value of $\sigma_*$ is determined so that the dimension of $\phi$ is continuous across this transition:
\beq
\sigma_*=d-2\Delta_\phi^{\rm SRI}\equiv 2-\eta_{\rm SRI}\,,
\eeq
using the usual definition of the $\eta$ critical exponent.

The above picture is considered standard since the foundational work of \cite{Fisher:1972zz,Sak}; it has been supported by theoretical studies \cite{Honkonen:1988fq,Honkonen:1990mr} and by Monte-Carlo simulations \cite{Blote}.

There is some ongoing debate about what precisely happens near $\sigma=\sigma_*$. Some recent studies \cite{Picco:2012ak, Blanchard:2012xv} claim observing deviations from the standard picture in this region, while others provide further evidence in its support \cite{Angelini, Defenu:2014bea}.\footnote{Some of this recent literature \cite{Angelini, Defenu:2014bea} is phrased in terms of an idea of an ``effective dimension" $D_{\rm eff}(\sigma)$ such that the LRI critical point in $d$ dimensions is supposed to be equivalent to the SRI critical point in $D_{\rm  eff}$ dimensions. We find this idea highly implausible. 
The relations among a handful of LRI and SRI critical exponents, which led the authors of \cite{Angelini, Defenu:2014bea} to postulate this notion, are most likely approximate (see also \cite{Blanchard:2012xv}). In our work the idea of the effective dimension is not going to play any role.} In this work we will be focussing on the region near the other phase boundary $\sigma=d/2$, i.e.~away from $\sigma=\sigma_*$. We will therefore not try to weigh in on this debate, except for a small comment in the discussion section.

{\bf Problem of conformal invariance}
\nopagebreak

As we will review in section \ref{sec:SRIconf}, the SRI critical point enjoys the property of conformal invariance.
The utility of this symmetry in 2d is long known, as it allows for an exact solution of the critical theory \cite{Belavin:1984vu} via a method called the conformal bootstrap.\footnote{In fact, the 2d SRI critical point is invariant under the Virasoro algebra, which is an infinite-dimensional extension of the global conformal algebra. In this paper, we will be working in general $d$ and by conformal invariance we will mean global, finite dimensional, conformal invariance, unless specified otherwise. In 2d this is sometimes referred to as M\"obius invariance.} In 3d, an exact solution is not yet known, but the conformal bootstrap has recently been used to get the world's most precise numerical determinations of the SRI critical exponents \cite{Simmons-Duffin:2015qma}.\footnote{For prior work see \cite{Rattazzi:2008pe,Rychkov:2011et,ElShowk:2012ht,ElShowk:2012hu,El-Showk:2014dwa,Kos:2014bka}. An alternative technique has been developed in \cite{Gliozzi:2013ysa,Gliozzi:2014jsa,Gliozzi:2015qsa}.}

What about the LRI critical point? In region 3 it's identical to the SRI so it's conformal. In region 1 it's described by the nonlocal gaussian theory, whose conformal invariance is well known and will be reviewed in section \ref{sec:gaussconf}. 

Our main goal here will be to show that the LRI critical point is conformal also in the non-gaussian region 2. 
This is harder to ascertain, and as far as we know, this issue has not been previously discussed. Here we will present a proof valid to all orders in perturbation theory. Having conformal symmetry also in this region is very interesting and useful, paving the way for the conformal bootstrap methods.\footnote{Some work has already been done in \cite{ElShowk:2012ht}, section 5.4, and \cite{Sheer}.} Until now, the LRI critical point was studied using RG methods, both perturbative \cite{Fisher:1972zz,Sak,Honkonen:1988fq,Honkonen:1990mr} and nonperturbative \cite{Defenu:2014bea}.

The remainder of this paper is organized as follows. In the next section we discuss the basic setup of the epsilon expansion for the long-range Ising model. We then provide nontrivial evidence for conformal invariance by computing $\vev{\phi \phi^3}$ and $\vev{\phi^2 \phi^4}$ up to order $\epsilon^2$, as we find that these correlators vanish at the fixed point. This behavior does not follow from scale invariance alone but it is a necessary condition for conformal invariance, so we are led to believe that the LRI at criticality could actually be conformally invariant. The remainder of the paper is dedicated to an all-orders proof of this claim. In sections \ref{sec:SRIconf} and \ref{sec:gaussconf} we review the proofs of conformal invariance of the SRI and of the gaussian theory (see below for a definition). This sets the stage for the proof of conformal invariance of the LRI, which we present in section \ref{sec:LRIconf}. In that section we also discuss the prospects for proving conformal invariance at the nonperturbative level. We end the paper with some concluding remarks, and several technicalities have been relegated to two appendices.

\section{Field-theoretical setup}
\label{sec:setup}

In this work we will study the LRI critical point for $1\le d < 4$, inside the non-gaussian region 2, close to the boundary separating it from the gaussian region 1, i.e.~for 
\beq
\sigma=(d+\eps)/2,\qquad 0<\eps\ll 1\,.\footnote{If $d$ is close to 4, we should assume a stronger condition $\eps\ll 4-d$, so that we stay closer to the boundary between the regions 1 and 2 than to that between 2 and 3. In the opposite case $\eps\gtrsim4-d$ the structure of perturbation theory is modified due to the presence of a weakly irrelevant operator $\del^2\phi$. See \cite{Honkonen:1988fq,Honkonen:1990mr} for a discussion.}
\eeq
The UV dimension of the $\phi$ field is then
\beq
\Delta_\phi =(d-\eps)/4\,,
\label{eq:phiUV}
\eeq
so that the quartic term $\phi^4$ is a weakly relevant perturbation. The IR fixed point is then accessible in perturbation theory. As is standard \cite{Honkonen:1988fq,Honkonen:1990mr}, we will setup a perturbative expansion using the analytic regularization scheme, where Feynman diagrams are considered analytic functions of $\eps$. This is convenient since it allows one to evaluate integrals without introducing an explicit UV cutoff. 

As mentioned in the introduction, the relevant action is
\beq
S=S_0+\frac{g_0}{4!}  \int {\rm d}^d x\, \phi^4\,,\qquad S_0=\frac{\calN_\sigma}2 \int {\rm d}^dx\, \phi\ \fl\phi\,. 
\label{eq:S}
\eeq
The coefficient $\calN_\sigma$, whose precise value is unimportant, will be fixed so that the free $\phi$ two point function is normalized to one:
\beq
\langle \phi(x) \phi(0)\rangle_{S_0}=|x|^{-2\Delta_\phi}\,.
\label{eq:2pt0}
\eeq

Notice that the analytic regularization scheme is mass independent. 
So we can avoid introducing the mass term $m^2 \phi^2$ into the action \reef{eq:S}. In other words, in this scheme the flow which leads to the fixed point has $m^2=0$ all along the flow. 

As usual in the field theoretical studies of critical phenomena, it will be extremely useful to view the theory \reef{eq:S} as regulating the theory with $\eps=0$, where the interaction is marginal. As $\eps\to0$, computations in theory \reef{eq:S} give poles in $\eps$. These poles can be removed order by order in perturbation theory by multiplicatively renormalizing the terms in the action.\footnote{Multiplicative renormalizability is most familiar in local quantum field theory, but it holds also for nonlocal theories of the kind we are considering here, with the standard proof \cite{Collins:1984xc}. For the validity of Weinberg's convergence theorem \cite{Weinberg:1959nj} (that if all subdiagrams are superficially convergent, then the diagram is convergent) in this context see the discussion in \cite{ticciati_quantum_1999}, p.600. For the structure of divergences in theories with propagators involving arbitrary powers of $x^2$ (studied in the context of analytic regularization), see \cite{zav'yalov} and the discussion in \cite{Honkonen:1988fq}.}

The gaussian term in \reef{eq:S} is not renormalized, because it's nonlocal, while the divergences are local. 
In other words, the theory under consideration does not have wavefunction renormalization, and the bare and the renormalized field $\phi$ coincide. In particular, the anomalous dimension of $\phi$ vanishes. This implies that, to all orders in perturbation theory, the dimension of $\phi$ at the IR fixed point is equal to its UV dimension \reef{eq:phiUV} \cite{Fisher:1972zz,Sak,Honkonen:1988fq}.

On the other hand, the coupling constant does require renormalization. The bare coupling $g_0$ is expressed in terms of the finite, renormalized, coupling $g$ as:
\beq
g_0= Z(g,\eps) g \mu^\eps\,,
\label{eq:gg0}
\eeq
where $\mu$ is the renormalization scale and $Z$ is the renormalization factor which starts with $1$ and contains an ascending series of poles in $\eps$:
\beq
Z(g,\eps)=1+\sum\nolimits_{k=1}^\infty {f_k(g)}{\eps^{-k}}\,.
\label{eq:Z}
\eeq
The coefficient $f_k(g)$ of the pole $\eps^{-k}$ is a power series in $g$ starting from $O(g^k)$; it gets contributions from all loop orders larger than or equal to $k$.

The $\beta$-function of the theory can be expressed in terms of the single pole coefficient:
\beq
\beta(g)\equiv {\del g}/{\del\log \mu}=-\eps g+g^2 f_1'(g)\,.
\eeq
The story is exceedingly similar to the $\eps$-expansion for the Wilson-Fisher (WF) fixed point (see e.g.~the books \cite{Kleinert:2001ax,Vasilev:2004yr}, or \cite{Brown:1979pq} for a concise treatment), with an extra simplification that there is no wavefunction renormalization.

Let us compute the $\beta$-function at the first nontrivial, one-loop, order. We need to determine the one-loop counterterm to the coupling. Consider the four point function of $\phi$, which up to the second order in $g$ is given by the sum of diagrams:
\beq
\includegraphics[scale=0.6]{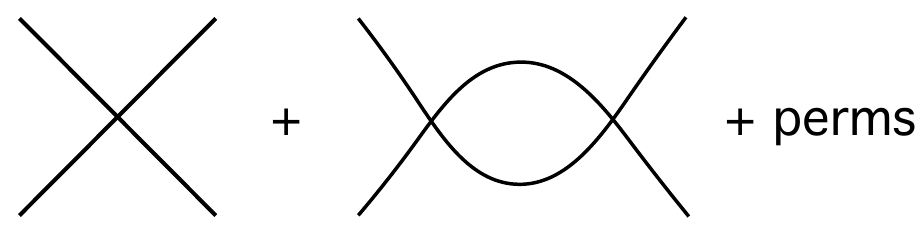}
\nn
\eeq
The $O(g^2)$ contribution will contain a $1/\eps$ pole coming from the region where the two $\phi^4$ insertions are near each other. The contribution of this region can be easily extracted using the operator product expansion (OPE) of the gaussian UV theory:
\beq
\phi^4(x) \times \phi^4(0)\supset ({72}/{|x|^{d-\eps}})\, \phi^4(0)\,.
\label{eq:ope4}
\eeq
This OPE implies that the expansion of the functional integral to the second order will contain a short-distance contribution corresponding to the insertion of $\phi^4$ times
\beq
72 \times \frac 12 \times \frac{g^2 \mu^{2\eps}}{(4!)^2} \int_{|x|\ll 1} \frac{\d^d x}{|x|^{d-\eps}} = 36 \frac{g^2}{(4!)^2}\frac {{\rm S}_d}{\eps} +\text{finite}\,,
\eeq
where ${\rm S}_d=2\pi^{d/2}/\Gamma(d/2)$ is the volume of the unit sphere in $d$ dimensions. The $1/\eps$ pole is canceled by adding a coupling counterterm
\beq
\delta g= (36 g^2/{4!}) ({\rm S}_d/{\eps})\,.
\eeq
Therefore the expansion of the single pole function $f_1(g)$ in \reef{eq:Z} starts with
\beq 
f_1(g)=K g+O(g^2),\qquad K=3{\rm S}_d/2\,.
\label{eq:K}
\eeq
The one-loop $\beta$-function is given by
\beq
\beta(g)=-\eps g+K g^2 +O(g^3)\,.
\eeq
It has a zero at 
\beq
g=g_*=\eps/K+O(\eps^2)\,.
\eeq
This zero corresponds to the LRI critical point that we want to study.\footnote{\label{note:Mitter} In $d=1,2,3$ and for sufficiently small $\eps>0$ the existence of this fixed point has been shown rigorously \cite{Brydges:2002wq,Abdesselam:2006qg,Mitter-talk,Mitter-email}. See also 
\cite{aizenmanCritical1988} for a series of rigorous results about the LRI phase transition. We are grateful to Abdelmalek Abdesselam and Pronob Mitter for communications concerning these works.}

Below we will be interested in the correlation functions of the composite operators $\phi^n$. As usual, we will define the renormalized operators $[\phi^n]=[\phi^n]_{g,\mu}$ which remain finite in the limit $\eps\to0$. They are related to the bare operators $\phi^n$ by rescaling factors subtracting poles in $\eps$:
\beq
\phi^n=Z_n(g,\eps) [\phi^n]\,.
\label{eq:[phin]}
\eeq
As indicated, the operators $[\phi^n]$ depend on the scale $\mu$ and on the value of the coupling $g$ at this scale. On the other hand correlation functions of the bare operators don't depend on $\mu$ and $g$ separately, but only on their combination $g_0$. As usual, this gives a CS equation for the correlators of the renormalized operators. The anomalous dimension of $[\phi^n]$ is given by
\beq
\gamma_n(g)=Z_n^{-1} {\del Z_n}/{\del \log\mu} |_{g_0=const}\,.
\eeq 
Notice that since the dimension of $\phi$ is not integer, in the considered theory the $\phi^n$ operators don't mix with the operators where some $\phi$'s are replaced by derivatives.\footnote{This is true for generic $d$, while for integer $d$ there can be some mixing for high $n$. E.g.~$\Delta_\phi\approx 1/2$ in $d=2$, and one can trade 4 $\phi$'s for 2 derivatives. This could become important for operators starting from $\phi^6$. In this paper we will mostly work with operators up to $\phi^4$ and we will ignore this.}

As already mentioned above, this theory has no wavefunction renormalization. Therefore $\phi$ is a finite operator without any rescaling. So $Z_1\equiv 1$ and $\gamma_1\equiv 0$. The UV dimension of $\phi$ given in \reef{eq:phiUV} will also be its IR dimension.\footnote{This fact has also been established non-perturbatively for sufficiently small $\eps>0$ for which the existence of the fixed point is rigorously known \cite{Mitter-talk,Mitter-email}, see note \ref{note:Mitter}.}

Let's compute the anomalous dimensions of $\phi^n$, $n\ge 2$, at the lowest nontrivial order. The $1/\eps$ contribution to the correlation functions of $\phi^n$ will come from the nearby $(g/4!)\phi^4$ insertions. In this region we can use the OPE generalizing \reef{eq:ope4}:
\beq
\phi^n(0)\times \phi^4(x)\supset [{6n(n-1)}/{|x|^{d-\eps}}] \phi^n(0)\,.
\eeq
Integration in $x$ gives a $1/\eps$ pole which must be canceled by rescaling the operator. We get
\beq
Z_n=1- K_n {g}/{\eps}+\ldots\,,\quad K_n={n(n-1){\rm S}_d}/{4}\ \Longrightarrow\ \gamma_n(g) = K_n g +O(g^2)\,.
\label{eq:Zn}
\eeq
At the IR fixed point we have
\beq
\gamma_n^*\approx K_n g_* = [n(n-1)/{6}]\eps +O(\eps^2)\,.
\label{eq:gamman}
\eeq
These are the same anomalous dimensions as in the usual $\eps$-expansion for the WF fixed point (see e.g.~\cite{Rychkov:2015naa}). Indeed, the answer at this order is controlled by the combinatorics of the Wick contractions in the gaussian UV fixed point, which is the same in the local free scalar theory and in the nonlocal theory defined by our $S_0$. Of course, it's not true that all computations are the same between the two theories. We will see some examples below.

\section{Tests of conformal invariance}
\label{sec:test}

How can we check if a certain model is conformally invariant or just scale invariant? The currently available data on the LRI amount to the anomalous dimensions of scaling operators, which are determined with RG methods and also measured on the lattice from two point functions. These data cannot distinguish between scale and conformal invariance, because both predict the same functional form for the critical two point function:
\beq
\langle \calO(x) \calO(0)\rangle\propto |x|^{-2\Delta_\calO}\,.
\eeq
Here $\calO$ is a generic scalar operator of dimension $\Delta_\calO$. 

One celebrated prediction of conformal invariance is the form of a three point function \cite{Polyakov:1970xd}, but this is harder to compute and to measure on the lattice than the two point function. We are not aware of any three point function data for the LRI critical point. 

An easier discriminating variable is the two point function of two different operators. Scale invariance predicts that
\beq
\langle \calO_1(x) \calO_2(0)\rangle= c_{12}|x|^{-\Delta_1 -\Delta_2}\,,
\eeq
while conformal invariance implies \cite{Polyakov:1970xd} the stronger constraint that $c_{12}=0$ unless $\Delta_1=\Delta_2$. To be precise, this conclusion is reached if both operators are so-called primary operators, i.e.~if they are not derivatives of other operators. If the $\calO_i$ are both of the form $(\del^2)^{n_i} \calO$ for the same operator $\calO$, they will of course have a nonzero two point function. This trivial case is easy to monitor since the scaling dimensions are different by an even integer.

To summarize, conformal invariance implies that any two scalar operators whose dimensions are not different by an even integer must have a zero two point function, while scale invariance would allow such two point functions to be nonzero.

The goal of this section will be to study the following two point functions of different scalars at the critical point of the LRI:
\beq
\langle \phi(x) \phi^3(0)\rangle \text{ and }\langle \phi^2(x) \phi^4(0)\rangle\,.
\eeq
Notice that we chose pairs of operators with the same parity under $\phi\to-\phi$.
If the IR fixed point is only scale invariant but not conformal, these correlators could be nonzero.

We will consider these correlators in the perturbative setup of the previous section. As we will see, a nontrivial check requires to go to at least the second order in the coupling constant $g$, or in the parameter $\eps$ parametrizing the deviation from marginality. At $O(\eps^2)$, scale invariance alone then allows for both of the above correlators to be nonzero.

As a matter of fact, it turns out that the first correlator is a bit special, as it involves $\phi$ whose anomalous dimension is identically zero, as well as $\phi^3$ which is related to $\phi$ via a ``nonlocal equation of motion" (nonlocal EOM).\footnote{We thank Riccardo Rattazzi who emphasized this to us.} Using these facts, we will give an all-order argument that the correlator $\langle \phi\,\phi^3\rangle$ vanishes at the IR fixed point. We will buttress the argument by an explicit second-order computation showing that the correlator vanishes at $O(\eps^2)$.

We do not know any analogous general argument for the correlator $\langle \phi^2 \phi^4\rangle$. This seems to be a truly generic correlator involving fields with unrelated anomalous dimensions. We will perform an explicit second-order computation for this correlator, finding that it also vanishes at $O(\eps^2)$. 

These computations lead us to believe that the LRI critical point is in fact conformally invariant, and to look for a general proof of this fact. A proof that we found will be discussed in the rest of the paper.

\subsection{\texorpdfstring{$\langle \phi\,\phi^3\rangle$}{< phi phi3 >}}
\label{sec:13}

By dimensional considerations, this correlator should behave in the IR as $C|x|^{-\Delta_\phi-\Delta_{\phi^3}}$. We recall that at the WF fixed point $C\ne 0$ because $\phi^3$ is not a primary but is related to $\phi$ by the equations of motion: $\phi^3\propto \del^2 \phi$. In the conformal field theory (CFT) language, $\phi^3$ is a descendant of the primary $\phi$. At the LRI fixed point, we know for sure that $\phi^3$ is not a descendant of $\phi$ since the dimensions don't match. So $C\ne 0$ would disprove conformal invariance, while $C=0$ would be evidence in its favor.

We will first show that $C=O(\eps^3)$ by an explicit computation, and next present an argument that $C=0$ to all orders in perturbation theory. 

\subsubsection{Explicit computation}

We will now present an explicit computation of the correlation function $\langle\phi\, \phi^3\rangle$ to the second order in the coupling. More precisely, we will consider the correlator
\beq
F(x,g,\mu)=\langle\phi(x) [\phi^3](0)\rangle\,,
\eeq
where $[\phi^3]=[\phi^3]_{g,\mu}$ is the $\phi^3$ rescaled by subtracting poles in $\eps$ as discussed in section \ref{sec:setup}.\footnote{Note that since we don't take the limit $\eps\to0$, operator renormalization is not strictly necessary. The rescaling factors $Z_n(g,\eps)$ are finite at finite $\eps$, and the $1/\eps$ terms in them never even become large, since $g=O(\eps)$ all along the RG flow. The role of renormalization is rather that of convenience, as it allows to cleanly separate the $O(1)$ effects coming from the powers of $g/\eps$ from the effects suppressed by powers of $g$ which are truly higher order. This makes the weakly coupled nature of the theory manifest.}
This correlator allows an expansion in powers of the renormalized coupling $g$ without poles in $\eps$. By dimensional reasons, it should have the form
\beq
F={f(s,g)}{|x|^{-\Delta_\phi-\Delta_{\phi^3}^{(0)}}}\,,
\label{eq:dimG}
\eeq
where $\Delta_{\phi^3}^{(0)}=3\Delta_\phi$ is the UV dimension of $\phi^3$ and $f$ is a function of dimensionless
variables $g$ and $s\equiv \mu|x|$.

We will have a Callan-Symanzik (CS) equation expressing the invariance of the theory under simultaneous changes of $g$ and $\mu$ leaving $g_0$ fixed:
\beq
[\mu \del_{\mu}+\beta(g)\del_g+\gamma_3(g)]F(x,g,\mu)=0\,.
\label{eq:CSphi3}
\eeq
The only difference from the usual $\phi^4$ theory is that the anomalous dimension of $\phi$ does not appear since it's identically zero. Substituting \reef{eq:dimG} into \reef{eq:CSphi3}, we get an equation for $f$:
\beq
[s\del_s+\beta(g)\del_g+\gamma_3(g)] f(s,g)=0\,.
\label{eq:CSf}
\eeq
At large distances $\beta(g)\to0$ and $\gamma_3(g)\to \gamma_3(g_*)$, and Eq.~\reef{eq:CSf} predicts
\beq
f(s,g)\approx C s^{-\gamma_3(g_*)}\,.
\label{eq:fIR}
\eeq
We thus rederived the result that at large distances the considered correlation function should go as $|x|^{-\Delta_\phi-\Delta_{\phi^3}}$ where $\Delta_{\phi^3}=\Delta_{\phi^3}^{(0)}+\gamma_3(g_*)$. This argument does not determine the value of $C$, which in particular may still turn out to be zero.

To determine the prefactor $C$, we have to match the CS equation to fixed-order perturbation theory. Recall that the full solution of Eq.~\reef{eq:CSf} can be written in the form:
\beq
f(s,g)=\hat f\bigl(\bar g(s,g)\bigr)\exp\left[-\int_1^s d\log s'\, \gamma_3\bigl(\bar g(s',g)\bigr)\right]\,,
\label{eq:CSfull}
\eeq
where $\bar g$ is the ``running coupling" solving the following differential equation with a boundary condition at $s=1$:
\beq
s\,\del_s \bar g=-\beta(\bar g)\,,\qquad \bar g|_{s=1}=g\,.
\eeq
In particular, at $s=1$ the function $f(s,g)$ reduces to $\hat f(g)$. Once this latter function is fixed from perturbation theory, the prefactor $C$ in \reef{eq:fIR} is found as
\beq
C=\hat f(g_*)\,.
\label{eq:Cf}
\eeq
We will now show that this vanishes up to order $\eps^2$.

For this we would like to extract $\hat f$ up to the second order in $g$. Consider first the nonrenormalized correlator 
\beq
F_0(x)=\langle\phi(x) \phi^3(0)\rangle\,,
\eeq 
which has no tree-level contribution. Up to the second order in the coupling it is given by the sum of two position-space diagrams:\footnote{The signs and combinatorial factors are left implicit.}
\beq
\includegraphics[scale=0.6]{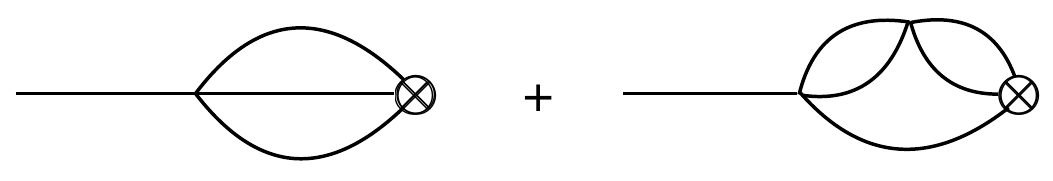}
\label{eq:fig-phiphi3}
\eeq
Both these diagrams are easily evaluated using the following basic integral (which in turn can be derived by going to momentum space):
\beq
\int \frac{{\rm d}^dy}{|x-y|^A|y|^B}=\frac {w_A\, w_B}{w_{A+B-d}} \frac{1}{|x|^{A+B-d}}\,,
\quad w_A=(4\pi)^{d/2}2^{-A}\frac{\Gamma\bigl(\frac {d-A}2\bigr)}{\Gamma(A/2)}\,.
\label{eq:basic}
\eeq
  Using this result, we obtain
  \beq
  F_0(x)= {R_1 g_0}/{|x|^{d-2\eps}}+ {R_2 g_0^2}/{|x|^{d-3\eps}} +O(g_0^3)\,,
  \eeq
  where
  \begin{align}
  R_1&=-\frac{w_\frac{d-\epsilon}{2} w_{3\frac{d-\epsilon}{2}}}{w_{d-2\epsilon}}\approx
  -\eps\, \pi^{d/2}\frac{\Gamma(-d/4)\Gamma(d/2)}{\Gamma(3d/4)}\,,\nn\\
  R_2&=\frac{3}{2}  \frac{w_{d-\epsilon}^2 w_\frac{d-\epsilon}{2} w_{\frac{3}{2}d - \frac{5}{2}\epsilon}}{w_{d-2\epsilon} w_{d-3\epsilon}}\approx 9\pi^d \frac{\Gamma(-d/4)}{\Gamma(3d/4)}\,.
  \label{eq:R1R2}
  \end{align}
The given approximate expressions are the leading ones in the small $\eps$ limit. Notice that $R_1=O(\eps)$. This has a simple explanation: in the limit $\eps\to0$ the interaction becomes exactly marginal, the integral defining $R_1$ becomes conformal, and it should give zero answer for a correlator of two fields of different scaling dimensions by the usual arguments based on conformal symmetry. To get a nontrivial check of conformal invariance, we have to go to the second order in $\eps$, hence to the second order in $g$, which is what we are doing.
  
To get at the function $\hat f$, we need to replace the coupling $g_0$ by $g$ via \reef{eq:gg0}, \reef{eq:K}:
\beq
g_0=Z g\mu^\eps=[g+Kg^2\eps^{-1}+O(g^3)]\mu^\eps\,,
\eeq
and to use the relation between $\phi^3$ and $[\phi^3]$, see \reef{eq:[phin]}, \reef{eq:Zn}. This gives the following expression for $F$ to the second order in $g$:
\beq
F(x)= \left\{g R_1 s^\eps + g^2 [R_1 (K+K_3)\eps^{-1} s^\eps + R_2 s^{2\eps}]\right\}/{|x|^{d-\eps}}\,.
\eeq
We see the structure is in agreement with \reef{eq:dimG}. To extract $\hat f$ we set $s=1$ and obtain:\footnote{We have analyzed also the leading $\log s$ terms in the expansion of $G$ around $s=1$ and checked that they are consistent with what the solution \reef{eq:CSfull} to the CS equation predicts.}
\beq
\hat f(g)=g R_1+g^2 [R_1 (K+K_3)\eps^{-1} + R_2]+O(g^3)\,.
\label{eq:fhat}
\eeq
Since $R_1=O(\eps)$ this expression is free of poles in $\eps$: as we mentioned above the correlator and in particular $\hat f$ should have a regular expansion in $g$ without such poles. 
Using \reef{eq:fhat} and \reef{eq:CSfull} we can compute the considered correlator at all distances with $\eps^2$ accuracy. 

Now using the values of the various constants appearing in \reef{eq:fhat}, it's easy to see that to the order that we computed it can be rewritten as
\beq
\hat f(g)=-{R_1}{\eps^{-1}} (-\eps g + K g^2) = -{R_1}{\eps^{-1}} \beta(g)\,,
\label{eq:fhat-rewrite}
\eeq
This form of writing makes it manifest that $\hat f(g)$ vanishes at the IR fixed point.

\subsubsection{General argument}

We will now give a general argument that $C=0$ to all orders in perturbation theory. The idea is to use the nonlocal EOM of the LRI field theory \reef{eq:S}:\footnote{Notice that it would be a non-permissible stretch of terminology to call $\phi^3$ a ``nonlocal descendant" of $\phi$ on the grounds of this equation. Descendants are defined as local derivatives of primaries, and we are not aware of any useful generalization of the descendant concept to nonlocal relations.}
\beq
\calN_\sigma \fl\phi+\frac{g_0}{3!}\phi^3=0 \,.
\label{eq:nleom}
\eeq
This can be used to express the correlators of $\phi^3$ in terms of those of $\phi$. In particular:
\beq
\langle \phi^3(x)\phi(0)\rangle\propto \fl\langle \phi(x) \phi(0)\rangle\,.
\label{eq:rel13}
\eeq
This equation is subtle to use, because the fractional Laplacian is a nonlocal operator. We will therefore proceed cautiously.

First of all we have to understand in some detail the correlator $\langle \phi(x) \phi(0)\rangle$. Since $\phi$ does not acquire an anomalous dimension, its two point function has the form
\beq
\langle \phi(x)\phi(0)\rangle = {\rho (x) }{|x|^{-2\Delta_\phi}}\,,
\eeq
where
\beq
\rho(x)=\hat \rho \bigl (\bar g(s,g)\bigr )\,.
\eeq
is a function of the running coupling $\bar g$ which we introduced in the previous section. The function $\hat \rho$ can be determined by matching to perturbation theory. We will only need to know its rough structural properties. 

Through $O(g^2)$ we have two diagrams:
\beq
\includegraphics[scale=0.6]{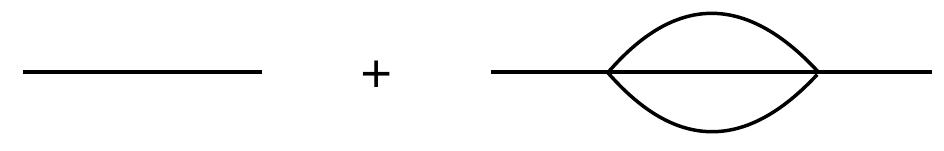}
\label{eq:ffdiags}
\eeq
This implies that 
\beq
\textstyle
\hat \rho(g)=1+ Q g^2+\ldots,\qquad Q=(\pi^d/6) {\Gamma\left(-\frac d4\right)}/{\Gamma\left(\frac{3d}{4}\right)}\,.
\label{eq:Q}
\eeq
We see that the function $\rho(x)$ approaches a constant at short and long distances:
\beq
\rho(x)\to \begin{cases} 1\,,& x\to 0\,,\\
\rho(g_*)=1+O(\eps^2)\,,& x\to\infty \,.
\end{cases}
\eeq

In what follows we will also need the asymptotics of the approach to the long distance limit. From now on we will fix $\mu$ to be a scale $\mu_c$ at which the coupling $g\sim g_*/2$ is roughly halfway between the UV and the IR fixed points.
The long distance asymptotic behavior of the running coupling $\bar g$ is given by:
\beq
\bar g = g_*- O(\eps s^{-\beta'(g_*)})\qquad (s \gg 1)\,,
\eeq
where $\beta'(g_*)=\eps+O(\eps^2)$. It follows that $\rho(x)$ approaches the long distance limit as\beq
\rho(x)= \rho(g_*)+O(\eps^2 s^{-\beta'(g_*)})\,.
\label{eq:hld}
\eeq

We are now ready to compute the long distance behavior of the correlator $\langle \phi^3(x) \phi(0)\rangle$. According to Eq.~\reef{eq:rel13} we need to evaluate the integral:
\beq
I=\int {\rm d}^d y\, T(x-y)\, \langle \phi(y)\phi(0)\rangle\,,
\eeq
where $T(x-y)\propto |x-y|^{-d-\sigma}$ is the position space kernel of the fractional Laplacian, see eq.~\reef{eq:kernel}. Notice that this kernel is not absolutely integrable and at short distances it must be understood in the sense of distributions, as the Fourier transform of $|k|^\sigma$.

We split the above integral into two parts $I=I_1+I_2$, one against the long distance asymptotics of $\langle \phi\, \phi\rangle$ and the rest:
\begin{align}
I_1&=\int {\rm d}^d y\, T(x-y) \rho(g_*)|y|^{-2\Delta_\phi}\,,\\
I_2&=I-I_1=\int {\rm d}^d y\, T(x-y) [\rho(y)-\rho(g_*)]|y|^{-2\Delta_\phi}\,.
\end{align}

The first part $I_1$ vanishes at non-coincident points, $I_1\propto \delta(x)$, by the definition of Green's function of the gaussian theory. 
As to $I_2$, in the limit of very large $x$ the leading asymptotics of the integral will come from large $y$, where we can use the asymptotics \reef{eq:hld}. We get:
\beq
I_2\sim \int {\rm d}^d y\, T(x-y) \eps^2 s^{-\beta'(g_*)} |y|^{-2\Delta_\phi}\,.
\eeq 
By dimensional analysis, this integral behaves $\propto|x|^{-\alpha}$ with
\beq
\alpha=\sigma+2\Delta_\phi+\beta'(g_*)\,.
\eeq
It's important that this exponent is larger than $\Delta_\phi+\Delta_{\phi^3}$:
\beq
\alpha-(\Delta_\phi+\Delta_{\phi^3})=\beta'(g_*)+O(\eps^2)\,,
\eeq
using the known anomalous dimension of $\phi^3$, $\gamma_3(g_*)=\eps+O(\eps^2)$.
This means that the constant $C$ in the natural dimensional asymptotics of the studied correlator, see \reef{eq:fIR}, has to vanish. The current argument establishes this fact to all orders in perturbation theory.

The above reasoning was made possible by two properties: the nonlocal EOM and the vanishing anomalous dimension of $\phi$. Before leaving this section, we will similarly prove one more interesting fact: that $\gamma_3^*=\eps$ to all orders in perturbation theory. In other words the leading order result in \reef{eq:gamman} is in fact exact for $n=3$. 

We will use the nonlocal EOM to prove the following relation between the IR dimensions of $\phi$ and $\phi^3$:
\beq
\Delta_{\phi^3}=\Delta_\phi+\sigma \qquad\text{(IR)}\,,
\label{eq:phi13drel}
\eeq
of which $\gamma_3^*=\eps$ is an immediate consequence. Eq.~\reef{eq:phi13drel} looks similar to the relation $\Delta_{\phi^3}=\Delta_\phi+2$ valid at the WF fixed point,
also a consequence of the corresponding EOM.
However, due to nonlocality, the proof is a bit more subtle for the LRI.

The idea is to use the nonlocal EOM twice, expressing $\langle \phi^3\phi^3\rangle$ in terms of
$\langle \phi \phi \rangle$. This relation takes the form:
\beq
\langle \phi^3 \phi^3 \rangle=36 g_0^{-2} \calN_\sigma^2 \fl \fl \langle \phi \phi \rangle_{\rm pert}\,,\qquad
\langle\phi \phi \rangle_{\rm pert}\equiv\langle\phi \phi \rangle-\langle \phi \phi \rangle_{S_0}\,.
\label{eq:eomtwice}
\eeq
with the two $\fl$'s acting on each of the arguments of $\langle \phi \phi \rangle_{\rm pert}$. The subtlety here is that we have to subtract the gaussian two point function. This is easy to understand in perturbation theory. $\langle\phi \phi \rangle_{\rm pert}$ is the sum of all diagrams in which each $\phi$ is connected to a vertex, like in the second diagram in \reef{eq:ffdiags}. When we act with $\fl$'s, the legs connecting $\phi$'s to the vertices get cancelled, and we reproduce all diagrams for $\langle \phi^3 \phi^3 \rangle$. Were we to keep $\langle \phi \phi \rangle_{S_0}$, we would get an extra nonlocal contribution which does not correspond to any diagram. This is in contrast to what happens when using the equation of motion in the local $\varphi^4$ theory. In that case the contribution from the unperturbed propagator is zero at noncoincident points and we don't have to worry about it.\footnote{One can also check that \reef{eq:eomtwice} has a smooth limit when $g_0\to0$, reducing to $\langle \phi^3 \phi^3 \rangle_{S_0}$ in this limit. This would not be the case were we to keep $\langle \phi \phi \rangle_{S_0}$.}

Let us proceed now to the proof of \reef{eq:phi13drel}. It will be convenient to work in momentum space. The asymptotics \reef{eq:hld} of the function $\rho(x)$ means that the Fourier transform of $\langle \phi\,\phi\rangle_{\rm pert}$ behaves at small momenta as:
\beq
\langle \phi(-k)\phi(k)\rangle_{\rm pert} \sim |k|^{-\sigma}[1+O(|k|^{\beta'(g_*)})]\,.
\eeq
It's important that the proportionality coefficient here is nonzero (we computed that it's $O(\eps^2)$).
When we act on this correlator once by the fractional Laplacian, multiplying by $|k|^\sigma$, the leading term gives 1, which is a delta-function in the position space, and the subleading term determines the long-distance asymptotics. We thus reproduce the above result that $\langle \phi^3\,\phi \rangle$ vanishes in the IR faster than its natural scaling. However, when we act by the fractional Laplacian twice, the leading result is nonanalytic $|k|^\sigma$ and it is this result which determines the long-distance asymptotics. We conclude that the two point function $\langle \phi^3\,\phi^3 \rangle$ behaves at long distances as $1/|x|^{d+\sigma}$, which can be expressed as Eq.~\reef{eq:phi13drel}\,.


\subsection{\texorpdfstring{$\langle \phi^2 \phi^4\rangle$}{< phi2 phi4 >}}

In the previous section we studied the correlator $\langle \phi\, \phi^3\rangle$ and found that it vanishes at the IR fixed point, consistently with conformal invariance. However, as we saw, the studied correlator was actually a bit special, since it involved two fields related by the nonlocal EOM of the LRI. One could wonder if the vanishing of this correlator is an accident. In this section we will study the correlator $\langle \phi^2\, \phi^4\rangle$, which as far as we can see is a truly generic correlator of our theory.
We will find, by an explicit computation at $O(\eps^2)$, that this correlator also vanishes at the IR fixed point. 
We consider this a strong piece of evidence for the conformal invariance of the LRI critical point.

The computation proceeds similarly to $\langle \phi\, \phi^3\rangle$, so we will be brief. The renormalized correlator
\beq
H(x,g,\mu)=\langle [\phi^2](x)\,[\phi^4](0)\rangle = h(s,g,\mu)|x|^{-\Delta_{\phi^2}^{(0)}-\Delta_{\phi^4}^{(0)}}\,
\eeq
satisfies a CS equation which can be solved to give:
\beq
h(s,g,\mu)=\hat h\bigl(\bar g(s,g)\bigr)\exp
\left\{-\int_1^s d\log s'\, \left[\gamma_2\bigl(\bar g(s',g)\bigr)+\gamma_4\bigl(\bar g(s',g)\bigr)\right]\right\}\,.
\eeq
The function $\hat h$ is determined by matching to perturbation theory. We have one diagram at $O(g_0)$
and three diagrams at $O(g_0^2)$:
\beq
\includegraphics[scale=0.6]{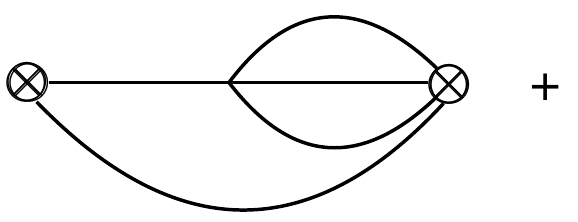} \quad
\includegraphics[scale=0.6]{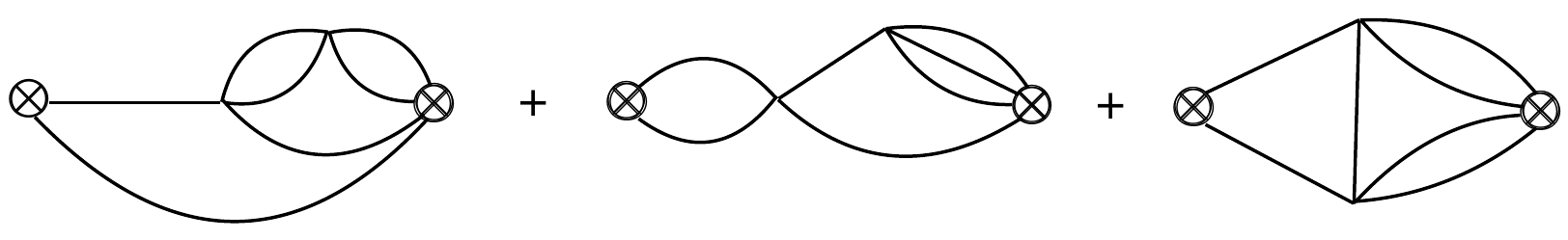}
\label{eq:phi2phi4diags}
\eeq
The first two diagrams here are the same as in \reef{eq:fig-phiphi3} times an extra propagator. The third diagram is new, but it can also be readily evaluated using Eq.~\reef{eq:basic} repeatedly. The final diagram is the hardest---it is analyzed in appendix \ref{eq:hard}.

The nonrenormalized correlator is thus given by
\beq
H_0(x)= \frac{P_1 g_0}{|x|^{3d/2-5\eps/2}}+ \frac{(P_{2a}+P_{2b}+P_{2c}) g_0^2}{|x|^{3d/2-7\eps/2}} +O(g_0^3)\,,
  \eeq
where we split the coefficients diagram by diagram. Taking into account signs and symmetry factors, we have (see \reef{eq:R1R2}):
\begin{align}
P_1&=8R_1,\nn\\
P_{2a}&=8R_2\,,\nn\\
P_{2b}&=4 \frac{ w_{\frac{3}{2}(d - \epsilon)} w_{\frac{d - \epsilon}{2}} w_{d - \epsilon} w_{\frac{3}{2}d - \frac{5}{2}\epsilon}}{w_{d-2\epsilon}w_{\frac{3}{2}d-\frac{7}{2}\epsilon}}\approx 8 \pi ^d \frac{\Gamma \left(-\frac{d}{4}\right)}{\Gamma
   \left(\frac{3 d}{4}\right)}\,,\nn\\
P_{2c}&\approx \textstyle  24 \pi ^d {\Gamma\left(-\frac d4\right)}/{\Gamma\left(\frac{3d}{4}\right)}\,.
   \label{eq:Pc}
\end{align}
To pass from this result to the function $\hat h$, we need to perform the coupling and operator renormalization. We get:
\beq
\hat h(g)=g P_1+g^2[P_1(K+K_2+K_4)\eps^{-1}+P_{2a}+P_{2b}+P_{2c}]+O(g^3)\,.
\eeq
Putting in the values of all coefficients, we find that this expression is proportional to the one-loop beta function, similarly to \reef{eq:fhat-rewrite}. We conclude that $\hat h(g)$ vanishes at the IR fixed point at $O(\eps^2)$.


\section{Conformal invariance of the SRI critical point}
\label{sec:SRIconf}

In the previous section we found that two point functions $\langle \phi\,\phi^3\rangle$ and $\langle \phi^2\,\phi^4\rangle$ vanish at the critical point of the LRI model. Nonzero values of these correlators would be consistent with the scale invariance and the $\bZ_2$ symmetry of the IR fixed point. We know of no symmetry but conformal invariance which could forbid these two correlators. We are led to believe\footnote{Contrary to the initial conviction of some of us.} that the model is conformally invariant, and we must understand why.

The phase diagram of the LRI model in Fig.~\ref{fig-standard} has three regions. Here we are interested in the conformal invariance inside the nontrivial region 2. In region 1 the critical point is gaussian, while it coincides with the SRI in region 3. The latter two cases are known to be conformally invariant, as will be reviewed in this and the next section. Understanding these arguments will be essential for the proof in the case of interest, in section \ref{sec:LRIconf}.

We will focus here on the proof of conformal invariance of the SRI critical point to all orders in the $\vareps$-expansion. 
Consider thus the usual, local $\varphi^4$ theory in $D=4-\vareps$ dimensions:\footnote{We will use $\varphi,D,\vareps$ in this section not to confuse with $\phi,d,\eps$ used in the sections dealing with the LRI. In the SRI case there is wavefunction renormalization, so we have to distinguish between the bare field $\varphi_0$ and the renormalized $\varphi$.}
\beq
S=\int {\rm d}^D x\left[ \frac 12 (\del\varphi_0)^2+\frac{\lambda_0}{4!}  \, \varphi_0^4\right]\,.\label{eq:phi4}
\eeq
The model is regularized and renormalized using the dimensional regularization and the minimal subtraction scheme.
There is an IR fixed point, called the Wilson-Fisher (WF) fixed point, which provides  a field-theoretic description for the SRI critical point.
We are specializing directly to the massless case, since in the considered scheme the mass renormalizes multiplicatively, and the fixed point is reached for zero bare mass. 

The idea of the proof \cite{Parisi:1972zy,Sarkar:1974xh,Brown:1979pq,Braun:2003rp} is to construct the Ward identities for scale and conformal invariance broken by the effects of the running coupling, and to show that the breaking effects disappear at the fixed point. We will follow here a simplified version of the presentation by Brown \cite{Brown:1979pq}.\footnote{Brown's goal was to construct a finite, renormalized, stress tensor, which he then also used to discuss the Ward identities. Since here we are only interested in the Ward identities, we can derive them using the canonical stress tensor, differing from Brown's finite tensor by total derivative terms. This allows us to avoid a complicated discussion of tensor operator renormalization, using only the rather simpler Eq.~\reef{eq:phi4Brown}.}

Consider the canonical stress tensor of the theory:
\beq
T_{\mu\nu}=\del_\mu\varphi_0\, \del_\nu\varphi_0-\delta_{\mu\nu}\left\{\half (\del\varphi_0)^2+(\lambda_0/4!) \, \varphi_0^4\right\}\,.
\eeq
It's easy to compute its divergence and trace:
\begin{gather}
\del^\nu T_{\mu\nu}=-E_\mu\,,\label{eq:div}\\
\qquad T^\mu{}_\mu = -\vareps \frac{\lambda_0}{4!}\varphi_0^4 -(D/2-1)E+ (1/2-D/4)  \del^2\varphi_0^2\,,
\label{eq:tr0}
\end{gather}
where the $E$ and $E_\mu$ are operators proportional to the equations of motion of the theory:
\begin{gather}
E=\varphi_0\left\{-\del^2\varphi_0+(\lambda_0/3!)\varphi_0^3\right\}\,,\qquad E_\mu=\del_\mu \varphi_0\left\{-\del^2\varphi_0+(\lambda_0/3!)\varphi_0^3\right\}\,.
\end{gather}
Correlation functions of these operators are trivial: their insertions into $N$-point functions of $\varphi$ produce a bunch of $\delta$-function at coincident points, see Brown's (3.10), (3.28):
\begin{gather}
G(x_1\ldots x_N; E(x))=\sum\nolimits_{i=1}^N \delta(x-x_i) G(x_1\ldots x_N)\,,\nn \\
G(x_1\ldots x_N; E_\mu(x))=\sum\nolimits_{i=1}^N \delta(x-x_i) \frac{\del}{\del x^\mu_i }G(x_1\ldots x_N)\,.
\label{eq:ins}
\end{gather}
One may wonder why we bother at all about the operators $E$ and $E_\mu$, since they have vanishing correlation functions at non-coincident points. Indeed, in the usual CFT language, they would not even qualify to be called operators. However, here we are working in perturbative quantum field theory, and in this situation it turns out to be both legitimate and useful to have access to $E$ and $E_\mu$. Legitimate because in the regularized theory the equations \reef{eq:ins} make perfect technical, and not just formal, sense. Useful because the fastest derivation of the Ward identities uses these operators, as we will see momentarily. 
 
 We will need next Brown's Eqs.~(3.13), (3.14) and (3.18) which relate the nonrenormalized operator $\varphi_0^4$ to the renormalized finite operator $[\varphi^4]$:\footnote{In accord with the rest of this paper, we denote by $\beta(\lambda)$ the full beta-function, equal to $-\vareps\lambda+\beta_{\rm Brown}(\lambda)$.}
 \beq
  \frac{\lambda_0}{4!}\varphi_0^4=-\frac{\mu^\vareps\lambda }{4!}(\beta(\lambda)/\vareps)[\varphi^4]+(\gamma(\lambda)/\vareps)E+ const.\del^2 \varphi_0^2\,.
  \label{eq:phi4Brown}
 \eeq
Using this relation in \reef{eq:tr0}, we get:
\beq
T^\mu{}_\mu = \frac{\mu^\vareps}{4!}\beta(\lambda)[\varphi^4] -\Delta(\lambda) E+ const'. \del^2\varphi_0^2\,
\,,\label{eq:tr}
\eeq
where $\Delta(\lambda)=(D-2)/2+\gamma(\lambda)$ is the full (classical + anomalous) dimension of $\varphi$. 

One could (as Brown does) construct an improved stress tensor whose trace would not contain a $\del^2\varphi_0^2$ term. This requires adding to $T_{\mu\nu}$ the well-known improvement term $\xi (\del_{\mu}\del_{\nu}-\eta_{\mu\nu}\del^2)\varphi_0^2$. At the classical level $\xi=-(D-2)/[4(D-1)]$, but it gets corrected at the quantum level \cite{Brown:1979pq}. If one is only interested in the Ward identities, as we are here, all this extra work can be avoided, since $\del^2 \varphi_0^2$ will drop out.

So, we proceed with the derivation of the Ward idenities. Consider the objects 
 \beq 
 \calD_\mu = T_{\mu\nu} x^\nu\,,\qquad \calC_\mu{}^\lambda = T_{\mu\nu} (2x^\nu x^\lambda-\delta^{\nu\lambda}x^2)\,.
 \eeq
 These are called scale and special conformal currents.
By \reef{eq:div} their divergences can be expressed as:
 \begin{gather}
 \del^\mu \calD_\mu = - x^\mu E_\mu+T^\mu{}_\mu\,,
 \label{eq:Ddiv}\\
 \del^\mu \calC_\mu{}^\lambda = - (2x^\mu x^\lambda-\delta^{\mu\lambda}x^2) E_\mu+2 x^\lambda T^\mu{}_\mu\,.
 \label{eq:Cdiv}
 \end{gather}
 Inserting these equation in an $N$-point function and integrating over the full space, we get:\footnote{The boundary term arising when integrating by parts vanishes. To show this one has to estimate a correlation function of a group of $\varphi$'s and a widely separated $T_{\mu\nu}$. In the critical theory and for a traceless $T_{\mu\nu}$ this would be $1/|x|^{2D}$ by using the stress tensor dimension. But here we need an estimate for a theory which has not yet reached the fixed point, and for an unimproved $T_{\mu\nu}$. For a quick and dirty estimate, notice that there will be at least two propagators connecting $T_{\mu\nu}$ to the $\varphi$'s, with two derivatives acting on them. This gives decay $\sim 1/|x|^{2D-2}$, which is more than sufficient for $D$ near 4. The $\varphi^4$ part of  $T_{\mu\nu}$ gives an even smaller contribution. An analogous argument justifies integrating by parts $\del^2\varphi_0^2$ a few lines below.}
 \begin{gather}
 \int \d^Dx \,G(x_1\ldots x_N; -x^\mu E_\mu(x)+T^\mu{}_\mu(x))=0\,,\\
 \int \d^Dx \,G(x_1\ldots x_N; - (2x^\mu x^\lambda-\delta^{\mu\lambda}x^2) E_\mu(x)+2 x^\lambda T^\mu{}_\mu(x))=0\,.\label{eq:SCT0}
 \end{gather}
 The contribution of $\del^2\varphi_0^2$ in $T^\mu{}_\mu$ vanishes after integrating by parts. Contributions of the other terms can be all expressed using \reef{eq:tr} and \reef{eq:ins}. We finally obtain: 
 \begin{gather}
 \sum _{i=1}^N [x_i.\del_{x_i} +\Delta(\lambda)]G(x_1\ldots x_N) = \beta(\lambda)\frac{\mu^{\vareps}}{4!} \int \d^Dx \,G(x_1\ldots x_N;[\varphi^4](x)) \,,
 \label{eq:dil}
 \\
 \sum _{i=1}^N \left[(2x_i^\mu x_i^\lambda-\delta^{\mu\lambda}x_i^2)\frac{\del}{\del{x^\mu _i}} +2\Delta(\lambda) x_i^\lambda \right]G(x_1\ldots x_N) =2 \beta(\lambda)\frac{\mu^{\vareps}}{4!} \int \d^Dx\, x^\lambda \,G(x_1\ldots x_N;[\varphi^4](x)) \,.
 \label{eq:SCT}
 \end{gather}
These are the Ward identities expressing the breaking of scale and special conformal invariance by the running coupling. 
They are valid at all distances and for all $\lambda$. For the purposes of studying the IR fixed point, we have to go to large distances and to take the limit $\mu\to0$, so that $\lambda(\mu)\to \lambda_*$. 
In this limit the beta-function multiplying the RHS vanishes. One is thus tempted to conclude that the IR fixed point is both scale and conformally invariance. 

While the conclusion is certainly correct, we believe that this last step of the argument merits somewhat more attention.
There is a subtlety here. It's true that $\beta(\lambda)\to0$ as $\mu\to0$, but what about the integrals multiplying it?
The integrals are $\mu$-dependent because the operator $[\varphi^4]$ is normalized at the scale $\mu$. Could it be that the integrals grow in the $\mu\to0$ limit, overcoming the $\beta(\lambda)\mu^\vareps$ suppression? This issue seems to have been neglected in the literature. In the next section we analyze the situation in detail and prove that this does not happen: the integrals behave as $O(\mu^{-\vareps})$, and so inferring the suppression from $\beta(\lambda)$ is correct. Our argument is a bit technical, and it would be nice to find a simpler proof.\footnote{\label{note:xdep}Such a proof can perhaps be given in the framework of the local CS equation \cite{Osborn:1991gm,Baume:2014rla} for the $\varphi^4$ theory with $x$-dependent $\lambda$ coupling. We thank Hugh Osborn for sharing with us his thoughts and preliminary results in this direction.}

Taking that argument for granted, we have shown that the long-distance limit of the correlation functions of $[\varphi]$ transforms covariantly under the conformal group, with $[\varphi]$ being a primary field of dimension $\Delta(\lambda_*)$. For a complete proof of conformal invariance of the SRI critical point, one needs to extend this result to correlators of composite operators. Correlation functions of composites can be related to those of the fundamental field by using the OPE, and it seems rather plausible that they will also be conformal. In appendix \ref{sec:ope} we sketch a proof of this fact.

At this point one may ask, analogously to what we did in section \ref{sec:test}: can conformal invariance of the SRI critical point be tested by checking that two point functions of different primaries vanish? 
As mentioned in section \ref{sec:13}, the correlator $\langle\varphi\, \varphi^3 \rangle$ is nonzero in this theory because $\varphi^3$ is a descendant of $\varphi$.
As for the $\langle \varphi^2\, \varphi^4\rangle$ correlator, we computed it and found that it does vanish at the IR fixed point to the first nontrivial order in $\vareps$, which in this theory is $O(\vareps)$.\footnote{The argument given below equation \reef{eq:R1R2} that $O(\eps)$ is trivial in the LRI does not apply in the SRI. The reason is that for $\vareps=0$ the SRI Feynman integrals are log-divergent. When moving to $\vareps>0$ they have $\vareps^{-1}$ poles enhancing the effect.} This computation is a bit nontrivial as it requires careful analysis of the mixing between the operators $\varphi^4$ and $\del^2\varphi^2$, so we don't present the details \cite{Rychkov:phi2phi4}.

\subsection{Scale and conformal invariance breaking effects}
\label{sec:Ward}

Here we will study carefully how the broken Ward identities \reef{eq:dil} and \reef{eq:SCT} imply scale and conformal symmetry of the WF fixed point. 

\subsubsection{A proof which is too quick, and a paradox}
\label{sec:morecareful}

Let us first discuss a deceptively quick proof.\footnote{See e.g.~p.29 of \cite{Braun:2003rp}.} Can't we just set $\lambda=\lambda_*$ in these equations? On the one hand, the correlators in the LHS then become the IR limits of the correlators of the full theory. On the other, the RHS side goes away as $\beta(\lambda_*)=0$. Right?

In our opinion, this argument is plagued with the following basic problem: the integrals multiplying $\beta(\lambda)$ become UV divergent for $\lambda=\lambda_*$. Indeed, for $y\to x_i $ we can use the OPE
\beq
\varphi(x_i)\times [\varphi^4](y)\sim \frac{const.}{|x_i-y|^{\Delta_{\varphi^4}}} \varphi(x_i),
\label{eq:ope14}
\eeq
where $\Delta_{\varphi^4}=D+\vareps+O(\vareps^2)$ is the IR dimension of $\varphi^4$. Thus the integrals diverge as $\Lambda^\vareps$, where $\Lambda$ is the UV cutoff marking the boundary where the asymptotics \eqref{eq:ope14} applies. In the strict $\lambda=\lambda_*$ limit we have $\Lambda=\infty$, the RHS of the Ward identities becomes $0\times\infty$, and we learn nothing. 


Clearly, we should resolve this $0\times\infty$ ambiguity by keeping $\lambda$ slightly different from $\lambda_*$. We should then identify $\Lambda$ with $\mu_c$--- the scale where the full RG flow transitions from the UV to the IR fixed point. This scale can be defined by the equation
\beq
\lambda(\mu_c)\sim \half \lambda_*\,.
\eeq
Since for $|x_i-y|\ll \mu_c^{-1}$ the dimension of $\varphi^4$ is near its UV fixed point dimension $D-\vareps$, the UV divergence of the integrals will be cut off at $\mu_c$. We then send $\mu_c\to\infty$, keeping $\mu$ and $x$ fixed. In this limit $\lambda$ approaches $\lambda_*$ as 
\beq
\lambda_*-\lambda \sim \vareps (\mu/\mu_c)^\vareps\,.
\eeq
We obtain
\beq
\beta(\lambda)\times(\text{integral})=O(\lambda_*-\lambda)\times \Lambda^\eps=O(\mu_c^{-\vareps}\times \mu_c^\vareps)=O(1)\,.
\eeq
Here we kept track only of the $\mu_c$ dependence, as other parameters are kept fixed. 

We thus have a paradox: in the $\mu_c\to\infty$ limit the suppression from the $\beta$-function seems to be exactly compensated by the growth of the integral. The RHS of the Ward identities seems to remain $O(1)$ and we show neither scale nor conformal invariance.

\subsubsection{Correct proof}

The resolution of the paradox is that the naive dimensional analysis does not capture the true asymptotics of the integrals multiplying the $\beta$-function. The estimate $O(\mu_c^\vareps)$ applies separately to the UV and IR parts of the integral, but as we will now explain, a cancelation occurs when the two are added. The full integral is then $O(1)$ instead of $O(\mu_c^\epsilon)$.

From now on, instead of taking $\mu_c\to\infty$ with $x,\mu$ fixed, it will be more convenient to keep $\mu_c$ fixed and approach the IR by taking $x\gg \mu_c^{-1}$, $\mu\ll\mu_c$. We will then have to keep track how various terms scale in the $\mu\to0$ limit. The basic issue---the insufficiency of the naive dimensional analysis---will however be the same as in the above paradox.

Starting with the scale Ward identity \reef{eq:dil}, we notice that it can in fact be analyzed in two ways.
One way is through the OPE as above:
\beq
\varphi(0)[\varphi^4](y)\sim C(\mu,y) \varphi(0)\,.
\eeq
The argument $\mu$ reminds that $[\varphi^4]$ is normalized at the scale $\mu$. 
In the $x\to\infty$ limit the RHS of \reef{eq:dil} can be approximated by using this OPE as
\beq
N\beta(\lambda) \frac{\mu^\vareps}{4!}\int \d^D y\,  C(\mu,y) \times G(x_1\ldots x_N)\,.
\label{eq:RHSscale}
\eeq
The second, actually more standard, way to analyze \reef{eq:dil} is rewrite its RHS as\footnote{See \cite{Brown:1979pq}, eq.~(3.19); factor $\mu^\eps$ is missing in the RHS of that equation.}
\beq
 \beta(\lambda)(\del/\del\lambda)G(x_1\ldots x_N)\,,
 \eeq
turning it into a CS equation, which can be solved similarly to the familiar two point function case. In particular this implies that at long distances the correlator is scale invariant with anomalous scaling dimensions:
 \beq
 G(\xi y_1,\ldots, \xi y_N)\propto \xi^{-N \Delta(\lambda_*)},\quad \xi\gg 1\,,
 \label{eq:scaleIR}
 \eeq
 where the points $y_1,\ldots y_N$ are assumed to have $O(\mu_c^{-1})$ distances among them.
 
Let us now take \reef{eq:scaleIR} and plug it into the LHS of \reef{eq:dil}. Assuming that all distances among $x_i$ are large compared to $\mu_c^{-1}$, the LHS of \reef{eq:dil} equals:
\beq
N(\Delta(\lambda)-\Delta(\lambda_*))G(x_1\ldots x_N)\,.
\eeq
Comparing this to the RHS as given in \reef{eq:RHSscale}, we get an interesting integral constraint on the OPE kernel:
\beq
\beta(\lambda)\frac{\mu^\vareps}{4!}\int \d^D y\,C(\mu,y)=\Delta(\lambda)-\Delta(\lambda_*)\,.
\eeq
Expanding this equation in $\lambda-\lambda_*$ for $\lambda(\mu)$ near $\lambda_*$ (i.e.~$\mu\ll\mu_c$), we get
\beq
\frac{\mu^\vareps}{4!}\times \int \d^D y\, C (\mu,y)=O(1) \qquad (\mu \ll \mu_c)\,.
\label{eq:intC}
\eeq
It's instructive to compare this result with simple dimensional analysis. 
By dimensional analysis, $C(\mu,y)$ must be of the form
\beq
C(\mu,y)=1/|y|^{D-\vareps}\times f(|y|\mu, \lambda)\,.
\eeq
Moreover, at large distances and to the leading order in epsilon it should scale $\propto 1/|y|^{D+\vareps}$, since the IR dimension of $[\varphi^4]$ is $D+\vareps+O(\vareps^2)$. We will ignore the admixture of $\del^2[\varphi^2]$ operator to $[\varphi^4]$, since such admixture does not contribute to the integrals in the RHS of the Ward identities.
We conclude that $f\propto (|y|\mu)^{-2\vareps}$ in the IR and thus
\beq
C(\mu,y)\propto 1 /[|y|^{D+\vareps}\mu^{2\vareps}],\qquad y\gg \mu_c^{-1}.
\label{eq:naive}
\eeq
Thus, were we to estimate the integral in \reef{eq:intC} by extending the integration from $\mu_c^{-1}$ to $\infty$, we would get a result which is not $O(1)$ but grows in the $\mu\to0$ limit as:
\beq
\mu^\vareps \times (\mu_c^{\vareps}/\mu^{2\vareps})=(\mu_c/\mu)^\vareps\,.
\eeq
This is the same growth which plagued the argument in section
\ref{sec:morecareful}. Apparently, there is a cancelation between the IR and the UV\footnote{The integral converges in the UV since the UV dimension of $\varphi^4$ is $D-\vareps$.} parts of \reef{eq:intC}. Separately they blow up in the $\mu\to0$ limit, but their sum remains $O(1)$.

To summarize the logic so far: there are two ways to analyse the scale Ward identity. The first is based on using the OPE which reduces to the integral of the OPE kernel. One needs further input to get a good estimate for this integral. This input comes from using the second way of analysis, which is based on the CS equation and gives scale invariance. From consistency of the two ways we obtain \reef{eq:intC}. 

Now to the special conformal Ward identity. Its RHS cannot be reproduced by differentiating the Lagrangian.\footnote{Unless one introduces $x$-dependent couplings, see note \ref{note:xdep}.} Thus there is only one way to proceed, that based on the OPE. We will now show, using \reef{eq:intC}, that the 
integral in the RHS of this Ward identity:
\beq
\mu^{\vareps} \int \d^Dy\, y^\lambda \,G(x_1\ldots x_N;[\varphi^4](y))\,
\eeq
remains for $\mu\to0$ of the same order of magnitude as the quantities appearing in the LHS. Multiplied by $\beta(\lambda)$ it vanishes, proving conformal invariance in the IR.

We will choose all $x_i$ and their distances large, of the order $s\gg\mu_c^{-1}$. We will also take $\mu\ll\mu_c$ to have the coupling closed to its fixed point value. Finally, we will assume $s\gg \mu^{-1}$. The rationale for this further separation of scales should become clear below.
We then divide the integral into three regions: 

1. \emph{$y$ is at distances $\lesssim \mu^{-1}$ from one of $x_i$.} Here we can use the OPE. We have $(z=y-x_i)$:
\beq
\mu^\vareps \int_{|z|\ll \mu^{-1}} C(\mu,z) (z^\lambda+x_i^\lambda)=0+x_i^\lambda \times O(1)
\eeq
where we used that $C(z)$ is spherically symmetric, and that the integral \reef{eq:intC} remains $O(1)$ even when we truncate it to $y\lesssim \mu^{-1}$. Indeed, the truncated away part of the integral can be estimated by naive dimensional analysis using \reef{eq:naive}, and it's $O(1)$.\footnote{One could ask about the contributions of the subleading OPE terms, like those proportional to $\del_\mu \phi$ and to the other operators. All such terms can be estimated using the naive dimensional analysis and they are small. Indeed, they are power suppressed compared to the leading term by at least one power of $y/s$. Since the naive dimensional analysis was only missing the true behavior of the leading term by a power of $\vareps$, it's more than sufficient for the subleading terms.}

2. \emph{$y$ at distances between $\mu^{-1}$ and $s$.} Here the OPE becomes less and less precise as the distance from $x_i$ increases. 
But we can still use it as an order of magnitude estimate, by saying that the insertion of $[\varphi
^4]$ suppresses the correlation function by (see eq.~\reef{eq:naive})
\beq
O(1/[|z|^{D+\eps}\mu^{2\vareps}]),\qquad \mu^{-1}\lesssim |z| \lesssim s\,.
\eeq
This is enough to conclude that the integral over this region is of the same order of magnitude as the LHS.

3. \emph{$y$ at distances $\gg s$ from the group of $x_i$.} The correlator decays as $1/|y|^{2\Delta_4}$. The integral is dominated by $y$ near the lower integration limit, which is the same as region 2.

This finishes the perturbative proof that the SRI correlations functions at the IR fixed point enjoy special conformal invariance.

\subsection[SRI conformal invariance beyond perturbation theory]{SRI conformal invariance beyond perturbation theory\protect\footnote{This section is outside of the main line of reasoning of this paper and can be left out on the first reading.}
}
\label{sec:SRIconfNP}

Let us briefly discuss what is known about the conformal invariance of the SRI critical point beyond perturbation theory. First of all, in 2d the SRI is exactly solvable on the lattice. Polyakov \cite{Polyakov:1970xd} first noticed that the critical three point correlation functions following from the exact solution satisfy the restrictions of global conformal symmetry. In the continuum limit the 2d SRI critical point is a minimal model CFT $\calM_{3,4}$ \cite{Belavin:1984vu}.  Aspects of gobal conformal and Virasoro invariance of the SRI critical point observables were also proven rigorously by mathematicians, pioneered by S.~Smirnov for nonlocal observables such as interfaces \cite{Smirnov-ICM}, 
and recently extended to correlation functions of local operators in bounded domains \cite{2dIsing-math-corr}. 

In 3d, the most dramatic evidence for the conformal invariance of the SRI critical point comes from its recent exploration using the conformal bootstrap methods \cite{Rattazzi:2008pe,Rychkov:2011et,ElShowk:2012ht,ElShowk:2012hu,El-Showk:2014dwa,Kos:2014bka,Simmons-Duffin:2015qma,Gliozzi:2013ysa,Gliozzi:2014jsa,Gliozzi:2015qsa}. These methods take conformal invariance as an assumption, which leads to a system of equations on the operator dimensions and OPE coefficients of the theory. The system is tightly constraining, and the fact that it has a solution which is in agreement with everything computed about the 3d Ising model using more pedestrian techniques is strong evidence for the validity of the assumption. 

Additional evidence for the conformal invariance of the 3d SRI critical point comes from the recent Monte-Carlo simulations of the model in the presence of a line defect \cite{Billo:2013jda} and in a ball geometry \cite{Cosme:2015cxa}.

Let us now discuss prospects for a nonperturbative \emph{proof} of the conformal invariance of the 3d SRI critical point. 
Because of the above evidence, we are 
certain that the model \emph{is} conformally invariant. 
Still, it would be interesting to find a proof, because it may teach us about the conformal invariance of other models where the evidence is as yet lacking.

Rather than deriving everything from scratch, it seems reasonable to allow a proof based on some assumptions. One rather minimal and natural set of assumptions
is as follows:\footnote{In the future it would be interesting to prove the 3d conformal invariance without any assumptions. 
For the moment we follow the advice of Ennio De Giorgi:
``Se non riesci a dimostrare l'enunciato, prendi una parte della tesi e mettila nelle ipotesi, e
continua cos\`i finch\'e non ci riesci."}
\begin{enumerate}
\item The 3d SRI critical point is a unitary quantum field theory, whose correlation functions satisfy the Wightman (or Osterwalder-Schrader in the Euclidean) axioms;
\item It is scale invariant, with all local operators having well-defined scaling dimensions;
\item It has a local stress tensor operator---a conserved rank 2 tensor operator of dimension exactly $D=3$, whose correlation functions satisfy Ward identities\,.
\end{enumerate}
Because of assumption 2, the problem is often formulated as ``Does scale invariance imply conformal invariance?"\footnote{See \cite{Nakayama:2013is} for a review and in particular \cite{Luty:2012ww,Fortin:2012hn,Dymarsky:2013pqa} for recent nontrivial progress in 4d.} As lucidly explained long ago by Polchinski \cite{Polchinski:1987dy}, this is not guaranteed in this setup. The crucial question is whether the trace of the stress tensor contains a total derivative term 
in addition to terms vanishing at the critical point or at non-coincident points:
\beq
T^\mu{}_\mu\supset \del_\mu V^\mu\,(?)\,
\label{eq:TK}
\eeq
The operator $V^\mu$ is called a virial current. It must be a vector operator of scaling dimension exactly $D-1$, to match the scaling dimension of $T_{\mu\nu}$. The appearance of $\del_\mu V^\mu$ in $T^\mu{}_\mu$ does not contradict scale invariance, because the extra term vanishes when integrated over the full space, and does not disturb the scale Ward identity. However, the derivation of special conformal Ward identity requires integrating $T^\mu{}_\mu$ multiplied by $x^\lambda$, see Eq.~\reef{eq:SCT0}. The extra term then does not go away \emph{unless} $V_\mu$ is itself a total derivative, $V_\mu=\del^\nu Y_{\mu\nu}$.\footnote{For example, the term $\del^2\varphi_0^2$ in $T^\mu{}_\mu$ did not disturb the derivation of the conformal Ward identity in section \ref{sec:SRIconf} because it corresponds to $V_\mu=
\del_\mu (\varphi_0^2)$, a total derivative.}

From a modern perspective, then, to check if a theory is conformally invariant it's enough to see if it contains a vector operator of dimension exactly $D-1$ which is (a) a singlet under all global symmetries, (b) not conserved, and (c) not a total derivative. In weakly coupled examples it can be seen by inspection whether such an operator exists. As Polchinski \cite{Polchinski:1987dy} pointed out, the $\varphi^4$ theory in $D=4-\vareps$ dimensions does not have any nontrivial $V_\mu$ candidate, and hence its IR fixed point must be conformally invariant to all orders in perturbation theory. 

In the strongly coupled situation one can argue 
(see e.g.~\cite{ElShowk:2011gz,EPFL,SNStalk}) that the (a,b,c) requirements cannot \emph{generically} be satisfied, and thus scale invariance \emph{generically} implies conformal invariance. Indeed, what's the likelihood that there will be a dimension $D-1$ vector which is not conserved, given that all non-conserved vectors generically acquire anomalous dimensions? This being pretty obvious, the real question is to what extent one can get rid of the genericity assumption.

Recently, the problem of showing conformal invariance of the 3d SRI critical point was revisited in \cite{Delamotte:2015aaa}. Using an argument phrased in the language of the functional renormalization group, the authors arrive at the same conclusion as Polchinski---that scale invariance without conformal invariance is only possible if the theory contains a dimension $D-1$ global singlet vector $V$. There is little doubt that their vector is the same as the virial current in Polchinski's argument, although this connection is not made explicit in \cite{Delamotte:2015aaa}.

The rest of the discussion in \cite{Delamotte:2015aaa} evolves around trying to show that a vector with dimension $\Delta_V=D-1$ is impossible. First they appeal to genericity. 
Compared to the above genericity argument, they add the following 
observation: to avoid conformal invariance, a dimension $D-1$ vector should exist in the WF fixed point not only for $D=3$, but in a whole interval of dimensions around 3, since otherwise one could argue conformality in $D=3$ by continuity from $D$ near 3. 

Second, they also point out that, independently of any theoretical arguments, the dimension of the lowest singlet vector can and should be measured in lattice Monte-Carlo simulations. If it's larger than $D-1$, as it probably is, this will imply conformal invariance. Such a test of conformal invariance is probably easier, and more robust, than the checks of its predictions carried out in \cite{Billo:2013jda,Cosme:2015cxa}.

To summarize, there is currently no nonperturbative proof of conformal invariance of the 3d SRI model critical point. Yet we are all but sure that it is conformally invariant.

{\bf Note added:} A third {\tt arXiv} version of \cite{Delamotte:2015aaa} just appeared which presents a proof of a lower bound $\Delta_V\ge D-1+\eta$, based on Lebowitz inequalities. If confirmed true, this would constitute a rigorous nonperturbative proof of conformal invariance in the 3d SRI.

\section{Conformal invariance of the gaussian phase}
\label{sec:gaussconf}

The nonlocal gaussian theory described by the action \reef{eq:kernel} is also conformal. This is widely known in high energy physics,\footnote{There, this theory is sometimes referred to as Mean Field Theory or Generalized Free Field.} and more recently has also been discussed from statistical physics perspective. In this section we will review this fact pedagogically, trying to bridge the gap between the two communities.

\subsection{Direct argument}

We will start in the antichronological order. As pointed out in \cite{Rajabpour:2011qr}, the fractional Laplacian, just like the ordinary Laplacian, is covariant under conformal transformations. Namely if $x\to x'$ is a conformal transformation and
\beq
\phi'(x')=|\del x'/\del x|^{-\Delta_\phi/d}\phi(x)\,
\eeq
with $\Delta_\phi$ given in \reef{eq:[phi]}, then
\beq
\fl'\phi'(x')=|\del x'/\del x|^{\Delta_\phi/d-1}\fl\phi(x)\,.
\label{eq:flinv}
\eeq
As a result the action \reef{eq:kernel} is invariant under conformal transformations. The proof of \reef{eq:flinv} is given in \cite{Rajabpour:2011qr} and we will not repeat it here. As usual, covariance under translations, rotations, and dilatations is obvious, and a simple calculation establishes covariance under the inversion.

Notice that since we are dealing with a nonlocal theory, there is no reason to expect that in $d=2$ the global conformal invariance of the action \reef{eq:kernel} gets enhanced to the full Virasoro invariance, and indeed this does not happen.\footnote{Although it's not essential for this paper, we would like to point out that there are also examples of \emph{local} 2d theories which have global conformal but not Virasoro invariance. Once such theory is the ``biharmonic scalar" with the Lagrangian $(\del^2\phi)^2$. Its stress tensor trace is of the form $\del^\mu\del^\nu Y_{\mu\nu}$ which is enough for global conformal invariance but not enough for an improvement to make it traceless and recover full Virasoro, since in 2d one needs $T_{\mu}{}^\mu=\del^2 Y$ for the latter \cite{Polchinski:1987dy}. See \cite{Wiese:1996xd,Rajabpour:2011qr} for a discussion.}

\subsection{Argument from correlation functions}

The previous argument shows the invariance of the action. Since the theory is gaussian, conformal covariance of the correlation functions follows. It's also easy to check the transformation properties of the correlation functions directly. This way of proving conformal invariance predates the previous one.
One of us (S.R.) was aware of it at the time of writing \cite{Rychkov:2009ij}, and an analogous discussion for a nonlocal \emph{vector} theory appeared in \cite{Dorigoni:2009ra}.

%

Let's start with the two point function of $\phi$. It is given by $|x-y|^{-2\Delta_\phi}$, which indeed has the form of a two point function of a primary scalar of dimension $\Delta_\phi$. The $N$-point functions of $\phi$ are given by Wick's theorem, since the theory is gaussian. A moment's thought shows that since the two point function transforms as it should, the $N$-point functions will do so as well. So all correlation functions of $\phi$ are consistent with conformal symmetry.

The theory contains more operators, for example the normal ordered products $\NO{\phi^n}$. Their correlation functions are defined by just leaving out the Wick contractions at coincident points, hence they will also be conformally covariant.

Although there are still more operators in addition to the ones considered above, all of them can be obtained by taking repeatedly the OPE of $\phi$ with itself. Correlation functions of these operators will inherit conformal transformation properties from the correlators of $\phi$.\footnote{See the argument in appendix \ref{sec:ope} for the WF fixed point.} Hence the theory is conformal.


\subsection{AdS argument}

Going further back in time \cite{Gubser:1998bc,Witten:1998qj}, an easy way to see conformal invariance of the theory \reef{eq:kernel} is from its dual holographic description. This description is given by the free massive scalar field $\Psi=\Psi(x,z)$:
\beq
S_{\rm AdS}=\half \int \d^dx\, \d z \sqrt{g} [g^{AB}\del_A \Psi \del_B \Psi  + M^2 \Psi^2]\,
\label{eq:Sholo}
\eeq
in the $d+1$ dimensional AdS space with the additional coordinate $z>0$ and the metric\footnote{This metric is considered fixed, rather than fluctuating like in the conventional AdS/CFT correspondence. This is related to the fact that the gaussian theory does not have a local stress tensor operator.}
\beq
ds^2=z^{-2}(dx^\mu dx_\mu + dz^2)\,. 
\eeq
 The dual theory \reef{eq:Sholo} is manifestly invariant under the conformal group $SO(d+1,1)$.
This is because AdS is a symmetric space, and $SO(d+1,1)$ is its group of isometries. Since the action \reef{eq:Sholo} is written using the generally covariant rules, isometries of the underlying space become symmetries of the action.

To show that the theory \reef{eq:Sholo} is indeed dual, i.e.~equivalent, to the nonlocal theory \reef{eq:kernel},
we must ``integrate out'' the bulk of the AdS space by solving the equations of motion of the field $\Psi(x,z)$,
and derive an effective action for its boundary value $\phi_b$ at $z=0$. The field $\Psi(x,z)$ behaves near the boundary as
\beq
\Psi(x,z)\sim z^{d-\Delta}\phi_b(x)\,,
\label{eq:bdrycond}
\eeq
where $\Delta$ is a root of the quadratic equation $\Delta(\Delta-d)=M^2$,
\beq
\Delta_\pm=\half (d\pm \sqrt{d^2+4M^2})\,.
\eeq
Picking $\Delta=\Delta_+$, the effective action has the form identical to \reef{eq:kernel} \cite{Gubser:1998bc,Witten:1998qj}:\footnote{This is the nonlocal part of the effective action. As is well known, the on-shell effective action contains divergent local counterterms which are implicitly subtracted here.
}
\beq
S_{\rm eff}\propto\int {\rm d}^dx\, \d^d y\, \phi_b(x)\phi_b(y)/|x-y|^{2\Delta}\,.
\label{eq:Seff}
\eeq
The range of dimensions $\Delta$ is however not quite the same. In fact $M^2$ should satisfy the so-called Breitenlohner-Freedman bound $M^2>-d^2/4$ \cite{Breitenlohner:1982bm}, which allows to reproduce all the dimensions above $d/2$, while in \reef{eq:kernel} we need dimensions in the range $(d-2)/2<\Delta<d/2$. As explained in \cite{Klebanov:1999tb}, dimensions in this range are reproduced by taking the smallest root $\Delta=\Delta_-$, which in this setup requires adding a boundary term in the action \reef{eq:Sholo}.

Since the original work of \cite{Gubser:1998bc,Witten:1998qj}, the above way of thinking about why the nonlocal gaussian scalar is conformal has been used numerous times in the high energy physics literature. See e.g.~\cite{Rychkov:2009ij,Heemskerk:2009pn,ElShowk:2011ag,Fitzpatrick:2011dm} for a partial list of recent references. This theory is very well understood from the CFT perspective, e.g.~in \cite{Heemskerk:2009pn,Fitzpatrick:2011dm} the conformal block decomposition and the $\phi\times\phi$ OPE have been worked out explicitly for any $\Delta$.

\subsection{Caffarelli-Silvestre trick}
\label{sec:CS}

The idea of the previous argument is to rewrite the nonlocal gaussian theory as an equivalent higher dimensional theory whose conformal invariance is manifest. We will now present a second way to implement this idea, based on an observation of Caffarelli and Silvestre \cite{Caffarelli}.

Consider a scalar field $\Phi=\Phi(x,y)$ where the extra coordinate $y$ now takes values in the flat Euclidean space of $p=2-\sigma$ dimensions. Hopefully the reader is not disturbed by the fact that $p$ is in general fractional.
In this space we consider the massless scalar field action:
\beq
S_{\rm CS}= \int \d^dx\, \d^p y\, [(\del_x \Phi)^2+(\del_y \Phi)^2]\,.
\label{eq:SCS}
\eeq
We will now show that this local action is equivalent to the nonlocal action \reef{eq:kernel}.

Let $\phi(x)$ be the value of the field $\Phi$ on the $d$ dimensional hyperplane $y=0$:
\beq
\phi(x)=\Phi(x,0)\,,
\label{eq:CSbd}
\eeq
and consider the effective action for $\phi$ obtained by integrating out the rest of the space. 

To find it we have to solve the equations of motion of the higher-dimensional theory subject to the boundary condition \reef{eq:CSbd}. It's clear that we can restrict to the sector of fields radially symmetric in the $y$ variable. In this sector the action becomes ($z=|y|>0$):
 \beq
S_{\rm CS}= {\rm S}_{p}\int \d^dx\, \d z\, z^{1-\sigma}[(\del_x \Phi)^2+(\del_z \Phi)^2]\,.
\label{eq:SCSrad}
\eeq
Recall that ${\rm S}_p$ is the unit sphere volume in $p$ dimensions. 
The equation of motion is 
\beq
\del_x^2 \Phi+(1-\sigma) \del_z\Phi/z +\del_z^2 \Phi=0\,.
\label{eq:eomCSrad}
\eeq
The solution 
decreasing at large $z$ takes in momentum space the form:
\beq
\Phi(p,z)=
const.(|p|z)^{\sigma/2} K_{\sigma/2}(|p|z)\, \phi(p)\,.
\eeq
For small $z$ this has an expansion:
\beq
\Phi(p,z)=(1+const.(|p|z)^{\sigma}+\ldots)\phi(p).
\eeq
The $\ldots$ terms are $O(z^2)$ compared to the shown ones and are subdominant in the range of interest $0<\sigma<2$.
We integrate by parts in \reef{eq:SCS}, pick up the boundary term, and find
\beq
S_{\rm CS}=-{\rm S}_{p}\int \d^d x\, z^{1-\sigma}\,\Phi\, \del_z \Phi|_{z\to 0}\propto \int \d^d p\, \phi(p)|p|^\sigma \phi(-p)\,,
 \eeq
 which is the nonlocal action \reef{eq:kernel} in momentum space.\footnote{The form of this effective action could also be predicted from the following argument. The free scalar $\Phi$ two point function depends on the distance as $\propto1/r^{d+p-2}$. The same dependence is inherited by $\phi$ when we set $y=0$, and we recover the expected two point function of the nonlocal scalar field of dimension $d-\sigma$.}

It is now easy to complete the argument for conformal invariance. We started with a massless scalar theory in flat space \reef{eq:SCS}, which is conformally invariant in an arbitrary number of dimensions. When we construct an effective theory by integrating out the space away from the $y=0$ hyperplane, we are guaranteed to obtain a theory invariant under the subgroup of the $d+p$ dimensional conformal group which leaves invariant this hyperplane. This is precisely the $d$ dimensional conformal group.

It's interesting to compare this argument with the AdS one in the previous section. 
The effective action computation here and in the AdS case proceed rather similarly. In fact the equations of motion of the AdS theory \reef{eq:Sholo} can be mapped onto those of \reef{eq:SCSrad} by defining 
$\Phi(x,z)=z^{-\Delta_\phi} \Psi(x,z)$. A minor difference is that the on-shell action is finite in the Caffarelli-Silvestre case, while in the AdS case it needs to be renormalized with boundary counterterms which we implicitly subtracted in \reef{eq:Seff}.

There is however a big difference in explaining why the equivalent higher-dimensional theory is conformally invariant. In the AdS case we appealed to the isometries of the underlying geometry, while in the CS description we use the fact that the massless scalar theory \emph{in flat space} is conformally invariant.

Concerning the prior history of the given argument, Caffarelli and Silvestre noticed that the equation of motion \reef{eq:eomCSrad} gives rise to the fractional Laplacian and showed the equivalence of the two actions. They also noticed that \reef{eq:eomCSrad} is nothing but the Laplace equations for functions radially symmetric in $p$ extra coordinates. The Caffarelli-Silvestre description in the radially reduced sector was used by Rajabpour \cite{Rajabpour:2011qr} to discuss some aspects of the nonlocal gaussian theory. Using the full $d+p$ dimensional description to argue for the conformal invariance of the theory seems to be done here for the first time.

\section{Conformal invariance of the LRI critical point}
\label{sec:LRIconf}

We will now present a proof that the LRI critical point is conformally invariant also in the nontrivial region 2.
After all the preparatory work in sections \ref{sec:SRIconf} and \ref{sec:gaussconf}, the proof looks almost inevitable.

The key idea is to rewrite our theory \reef{eq:S} as a \emph{defect quantum field theory}. Namely, according to the discussion in section \ref{sec:CS}, we can rewrite the action as\footnote{Normalization of the first term is different from \reef{eq:S} where it was fixed via \reef{eq:2pt0}. This difference is unimportant for the proof of conformal invariance.}
\beq
S= \half \int \d^{\bar d}X\, (\del_M\Phi)^2 + \frac{g_0}{4!} \int_{y=0} \d^d x\, \Phi^4\,.
\label{eq:S-rew}
\eeq
In the first term $\Phi$ is the free scalar Caffarelli-Silvestre field from section \ref{sec:CS}, defined on the $\bar d=d+p$ dimensional space $X=(x,y)$. The second term represents interaction living on the \emph{defect}: the $d$-dimensional plane located at $y=0$. The number of extra dimensions will be fractional in the case of interest ($\sigma$ near $d/2$), but in spite of this fact the theory \reef{eq:S-rew} makes perfect sense, at least in perturbation theory.\footnote{See section \ref{sec:LRI-NP} for a nonperturbative discussion. An introduction to perturbative quantum field theory in non-integer dimensions can be found in \cite{Collins:1984xc}.}

In the theory \reef{eq:S-rew}  we will consider $N$-point functions of $\Phi(X)$, $G(X_1\ldots X_N)$. The correlators of the original theory \reef{eq:S} can be obtained by taking the $y\to0$ limit.

The first part of the proof will be to derive broken scale and conformal Ward identities satisfied by the correlators $G(X_1\ldots X_N)$. We will then discuss how these Ward identities imply conformal invariance of the IR fixed point.

\subsection{Ward identities}

Crucially, the theory \reef{eq:S-rew} is local and so its Ward identities can be derived by a variation of the method from section \ref{sec:SRIconf}. We start by considering the canonical stress tensor:
\beq
T_{MN}=\del_M\Phi \del_N\Phi-\half \delta_{MN}(\del_K\Phi)^2 -\delta^\parallel_{MN} \delta^{(p)}(y) \frac{g_0}{4!} \Phi^4\,.
\eeq
The indices $M,N\ldots$ will run over the full $\bar d$-dimensional space, $\mu,\nu\ldots$ over the $d$-dimensional ``parallel" subspace, and $m,n\ldots$ over the $p$-dimensional ``perpendicular" subspace. The $\delta^\parallel_{MN}$ is the Kronecker delta in the parallel subspace: $\delta^\parallel_{MN}=\delta_{\mu\nu}$ if both indices are parellel, and zero otherwise. 

Unlike in section \ref{sec:SRIconf}, here we don't make a distinction between the bare and renormalized field $\Phi$, because $\Phi$ does not get renormalized: 
since the interaction term is located on the defect, it cannot renormalize the bulk kinetic term. The boundary kinetic term corresponds to an irrelevant operator and cannot be generated either.

The next step is to find the divergence and the trace of $T_{MN}$. Direct computation gives:
\begin{gather}
\del^M T_{MN}=-E_N+\delta^{(p)}(y) D_N\,,\\
T^M{}_{M}=-\Delta_\phi E -\eps ({g_0}/{4!}) \delta^{(p)}(y) \Phi^4 + (1/2-\bar d/4)\del_K^2 \Phi^2\,,
\label{eq:TMM}
\end{gather}
where we recognize the operators proportional to the equations of motion:
\begin{gather}
E= \Phi  \left\{-\del_K^2\Phi +\delta^{(p)}(y)(g_0/3!)\Phi ^3\right\}\,,\qquad
E_N=\del_N \Phi  \left\{-\del_K^2\Phi +\delta^{(p)}(y)(g_0/3!)\Phi ^3\right\}\,,
\end{gather}
and a new object $D_N$ called the displacement operator. It is given by: 
\beq
D_N=({g_0}/{3!}) \Phi ^3 \del_n \Phi \quad\text{if}\quad N=n\,
\eeq
is a perpendicular index, and vanishes otherwise. This operator represents an infinitesimal movement of the defect in an orthogonal direction.

Unlike $\Phi$, the operator $\Phi^4|_{y=0}\equiv \phi^4$ appearing in \reef{eq:TMM} is not a finite operator; it will be related to a finite renormalized operator $[\phi^4]$ via a rescaling
\beq
\phi^4 = Z_4 [\phi^4]\,.
\eeq
Contrary to \reef{eq:phi4Brown}, in the LRI there is no other operator with which $\phi^4$ could mix.
The coefficient $Z_4$ is easy to compute. For any finite correlation function we have
\beq
\begin{split}
\frac{\del}{\del g}\Bigr|_{\mu=const} \vev{\ldots} &= \vev{\ldots \int \d^d x \,\mu^\eps [\phi^4](x) }\,,\\
\frac{d }{d g_0} \vev{\ldots} &= \vev{\ldots \int \d^d x \, \phi^4(x) }\,.
\end{split}
\eeq
On the other hand, differentiating \reef{eq:gg0} with respect to $\log\mu$ we obtain the relation $\del g_0/\del g=-\eps g_0/\beta(g)$. It follows that:
\beq
Z_4=-\beta(g)\mu^\eps/(\eps g_0)\,,
\eeq
which is identical to the first coefficient in \reef{eq:phi4Brown}. Plugging this into \reef{eq:TMM} we obtain:
\beq
T^M{}_{M}=-\Delta_\phi E +(\beta(g)/{4!}) \mu^\eps \delta^{(p)}(y) [\phi^4]+ (1/2-\bar d/4)\del_K^2 \Phi^2\,.
\label{eq:TMM1}
\eeq
  
 We next consider the dilatation and special conformal currents:
\beq
 \calD_M = T_{MN} X^N\,,\quad
 \calC_M{}^L = T_{MN} (2X^NX^L-\delta^{NL}X^2)\,.
 \eeq
The divergence of the scale current is given by
\beq
 \del^M \calD_M = - X^N (E_N+\delta^{(p)}(y) D_N)+T^M{}_M= - X^M E_M+T^M{}_{M}\,.
 \eeq
 The term proportional to $D_N$ is seen to vanish, since either $N$ is a parallel index and then $D_N=0$, or else it's a perpendicular index and then $X^N \delta^{(p)}(y)=0$. The divergence equation takes the same form as \reef{eq:Ddiv}, and we obtain the scale Ward identity in full analogy to \reef{eq:dil}:
 \beq
 \sum _{i=1}^N [X_i.\del_{X_i} +\Delta_\phi] G(X_1\ldots X_N) = \beta(g)\frac{\mu^{\eps}}{4!} \int \d^d x \,G(X_1\ldots X_N;[\phi^4](x)) \,.
 \label{eq:scaleLRI}
 \eeq
 
Analogously let's analyze the divergence of the special conformal current 
\beq
 \del^M \calC_M{}^L = - (2X^N X^L-\delta^{NL}X^2) (E_N+\delta^{(p)}(y) D_N)+2 X^L T^M{}_M\,.
 \eeq
We would like to derive a Ward identity corresponding to special conformal transformations leaving the defect invariant,
i.e.~when $L=\lambda$ is a parallel index. The extra term proportional to $D_N$ then drops out just as for the scale current, since $D_N$ is nontrivial only if $N=n$ is a perpendicular index. Then $\delta^{n\lambda}=0$, while $X^n X^\lambda$ vanishes when multiplied by $\delta^{(p)}(y)$. We therefore have the same divergence equation as \reef{eq:Cdiv}, and an analogous Ward identity:\footnote{It's not hard to convince oneself, like we did in section \ref{sec:SRIconf}, that various boundary terms appearing when integrating by parts in the derivation of the Ward identities vanish by a good margin.}
\begin{multline}
 \sum _{i=1}^N \left[(2X_i^M X_i^\lambda-\delta^{M\lambda}X_i^2)\frac{\del}{\del{X^M _i}} +2\Delta_\phi X_i^\lambda\right]G(X_1\ldots X_N)\\
  =2 \beta(g)\frac{\mu^{\eps}}{4!} \int \d^d x\, x^\lambda \,G(X_1\ldots X_N;[\phi^4](x)) \,.
 \label{eq:scLRI}
 \end{multline}
 
Now, the correlators $G(X_1\ldots X_n)$ behave continuously in the limit $y\to0$. This can be seen diagram by diagram in perturbation theory. It's also closely related to the fact that $\Phi$ does not acquire anomalous dimension, and thus its bulk-to-defect OPE is non-singular (see a related discussion in section \ref{sec:btod} below).

Thus, if one is primarily interested in the correlators at the defect, which are the correlators of the original theory, then one may set $y=0$ in \reef{eq:scaleLRI} and \reef{eq:scLRI}, obtaining restricted Ward identities satisfied by the correlators of the original theory \reef{eq:S}: 
\begin{gather}
 \sum _{i=1}^N [x_i.\del_{x_i} +\Delta_\phi] G(x_1\ldots x_N) = \beta(g)\frac{\mu^{\eps}}{4!} \int \d^d x \,G(x_1\ldots x_N;[\phi^4](x)) \,,
 \label{eq:scaleLRI1}\\
\sum _{i=1}^N \left[(2x_i^\mu x_i^\lambda-\delta^{\mu\lambda}x_i^2)\frac{\del}{\del{x^\mu_i}} +2\Delta_\phi x_i^\lambda\right]G(x_1\ldots x_N) =2 \beta(g)\frac{\mu^{\eps}}{4!} \int \d^d x\, x^\lambda \,G(x_1\ldots x_N;[\phi^4](x)) \,.
 \label{eq:scLRI1}
 \end{gather}
At this point, if one wishes, one may altogether forget about the construct of the extra dimensions. Notice however that without this construct, it would remain pretty mysterious why the nonlocal LRI theory should satisfy Ward identities which are almost identical in form to the ones satisfied by the local SRI.\footnote{It is possible that the Ward identities \reef{eq:scaleLRI1}, \reef{eq:scLRI1} can be proved by endowing the LRI theory with some sort of nonlocal stress tensor operator. See \cite{Rajabpour:2011qr} for steps in this direction for the gaussian case. Still, the most natural path to the nonlocal stress tensor lies through the Caffarelli-Silvestre construction.} 
 
 \subsection{From the Ward identities to conformal invariance}
 
We will now discuss how the broken Ward identities imply that the IR fixed point is scale and conformally invariant. Naively this follows from the fact that $\beta(g)$ multiplying the breaking terms vanishes at the fixed point. However, as mentioned in section \ref{sec:SRIconf} for the SRI case, this is too quick. A careful argument must show that the integrals multiplying $\beta(g)$ do not overcome the suppression. For the SRI, this was done in detail in section \ref{sec:Ward}, and a modification of that argument will work for the LRI.

Let us first consider the $y=0$ correlators. The required modifications are then minor. The proof proceeds via the OPE $\phi \times [\phi^4]\supset \phi$. 
Using the scale Ward identity \reef{eq:scaleLRI1}, rewritten as a CS equation, one derives scale invariance,  and also obtains an integral constraint on the OPE kernel. There is a simplification due to the fact that $\Delta_\phi$ does not depend on $g$ in the LRI. Thus the LRI analogue of Eq.~\reef{eq:intC} will have 0 rather than $O(1)$ in the RHS. One then uses the integral constraint to show that the RHS of the special conformal Ward identity \reef{eq:scLRI1} is small in the IR. This part of the argument is identical to that for the SRI.

As for the correlators $G(X_1\ldots X_N)$ with fields inserted away from the defect, it seems a bit awkward to generalize the SRI argument to this case.\footnote{One would have to consider an OPE of the $[\phi^4]$ operator on the defect with $\Phi(x,y)$ in the bulk. The cases of $y$ small or large compared to $\mu_c^{-1}$ and the distances among $x_i$ appear to require separate treatments.} Fortunately there is an easier alternative argument: conformal invariance of these correlators can be seen to follow from two facts: (1) that for $y>0$ they satisfy the (conformally invariant) Laplace equation, and (2) that their boundary values at $y=0$ are conformally invariant, as we already proved.\footnote{This argument should be applied to the connected part of the bulk correlators, which satisfy the Laplace equation in each variable $X_i$ independently, without contact terms in the bulk.}

Furthermore, conformal invariance of the \emph{composite} operator correlators should now follow from by considerations based on the OPE. This is especially obvious for the bulk correlators, since one then needs only the bulk-to-bulk OPE, identical to that of the free theory. Composite operator correlators \emph{on the defect} can be obtained from those away from the defect via the bulk-to defect OPE (see the next section). Alternatively, one can access such correlators via the OPE of fundamental fields on the defect. This last way of arguing would be completely analogous to the one for the SRI given in appendix \ref{sec:ope}.

\subsection{Bulk-to-defect OPE and the anomalous dimensions}
 \label{sec:btod}
Up to now we were using the theory \reef{eq:S-rew} as a formal vehicle for proving things about the LRI. We would like to demonstrate, by a simple example, that this formulation can also be used for concrete computations.

One additional concept which will be necessary in this regard is the bulk-to-defect OPE.\footnote{For a similar concept of bulk-to-boundary OPE in boundary CFTs, see for example \cite{McAvity:1995zd,Liendo:2012hy}.} Let $\op (x,y)$ be a dimension $\D$ bulk operator of the theory. The bulk-to-defect OPE says that inside correlation functions insertions of this operator can be replaced by an infinite series of operators $\op_{k}$ inserted at the defect point $x$. By dimensional analysis, this expansion will take the form
\be
\op (x,y) = \sum_k \frac{f^k(g_0 y^\epsilon)}{|y|^{\D - \D_k^{(0)}}} \op_{k}^{(0)}(x)\,,
\ee
where the index $(0)$ marks bare defect operators and their classical dimensions. For simplicity we only consider scalar operators and we will ignore operator mixing. The renormalized bulk-to-defect OPE takes the form
\be
\op(x,y) = \sum_k \frac{\tilde f^k (y \mu,g)}{|y|^{\D - \D_k^{(0)}}} [\op_{k}](x)\,,
\ee
with the renormalized defect operators defined as
\be
\op_{k}^{(0)} = Z_k [\op_{k}]\,,
\ee
whereas the bulk operators do not renormalize. Clearly, then
\be
\mu \frac{d}{d \mu} \tilde f^k(y \mu, g) = \gamma_k(g) \tilde f^k(y \mu, g)\,,
\ee
which ties the anomalous dimension of the defect operators to the bulk-to-defect OPE. At the IR fixed point the bulk-to-defect OPE will take the form:
\be
\op(x,y) = \sum_k \frac{c_k}{|y|^{\D - \D_k^{\rm{IR}}}} [\op_{k}](x)\,,
\ee
This OPE will govern the asymptotics of any correlation function of $\op(x,y)$ in the limit when $y$ is much smaller than the $x$ distance between the operators.

As an example we consider the two point function of $\Phi^n$, with both operators inserted away from the defect. At $O(g)$ this correlation function is given by
\begin{gather}
\vev{\Phi^n(X_1) \Phi^n(X_2)} = \frac{n!}{|X_1-X_2|^{n\Delta_\phi}}\left\{1-
g \frac{n(n-1)}{4} (x_{12}^2 + y_{12}^2)^2  I[x_{12},y_1,y_2]\right\}\,,
\label{eq:oneloopex}\\
I[x_{12},y_1,y_2]= \int \d^d z [(z-x_{12})^2 + y_1^2]^{-2 \Delta_\phi} [z^2 + y_2^2]^{-2 \Delta_\phi} \,.
\end{gather}
The $I$ integral is finite for $y_{1,2}>0$. We can evaluate it using Feynman parameters:
\be
I[x_{12},y_1,y_2] = \frac{\pi^{d/2}\Gamma(4 \D_\phi - d/2)}{\Gamma(2\D_\phi)^2} \int_0^1 du \, [u(1-u)]^{2 \D_\phi - 1} \left[ u (1-u) x^2 + u y_1^2 + (1-u)y_2^2\right]^{d/2 - 4 \D_\phi}\,.
\ee
To simplify, let's consider the special case $y_1 = y_2 = y$. We Taylor expand in $x^2$, perform the $u$ integral for each term, and then resum the terms in the Taylor series. This leads to
\be
I[x_{12},y,y] = \frac{\pi ^{d/2} \Gamma \left(4 \Delta_\phi -\frac{d}{2}\right)}{\Gamma (4 \Delta_\phi )} y^{d-8 \Delta_\phi } \, _2F_1\left(2 \Delta_\phi ,4 \Delta_\phi -\frac{d}{2};2 \Delta_\phi +\frac{1}{2};-\frac{x_{12}^2}{4 y^2}\right)\,.
\ee
We are interested in the limit $y \ll x_{12}$. Because everything is finite we may set $\e = 0$, so $\D_\phi = d/4$.  For small $y$ we find then that
\be
I[x_{12},y,y] = - \Sd \frac{1}{x_{12}^d} [\gamma + \psi(d/2) + \log(y^2/x_{12}^2)] + O(y^2)\,.
\label{eq:Iappr}
\ee
We can substitute this result in the full correlation function and compare this with the behavior expected from the bulk-to-defect OPE:
\be
\vev{\Phi^n(x_1,y) \Phi^n(x_2,y)} \underset{y \ll x_{12}}{\to} \frac{C_n}{y^{2 \D - 2\hat \D} x_{12}^{2 \hat \D}}\,.
\label{eq:expected0}
\ee
Here $\D = n \Delta_\phi$, while $\hat \D$ is the dimension of the first defect operator, i.e.~$[\phi^n]$. To the leading order in $\eps$ that we are working, \reef{eq:Iappr} should match \reef{eq:expected0} expanded in the limit of small anomalous dimension $\gamma_n=\hat \D-\D$. We also have to set $g\to g_*$ in \reef{eq:oneloopex}. We obtain:
\be
C_n= n!\left[ 1 + g_*  K_n(\gamma + \psi(d/2)) + O(\eps^2)\right], \qquad \gamma_n = g_* K_n + O(\eps^2)\,,
\ee
where $K_n$ is the same as in \reef{eq:Zn}. The anomalous dimension result agrees with \reef{eq:gamman}. It is curious to see the usually scheme-dependent terms $\gamma$ and $\psi(d/2)$ appear in the squared bulk-to-defect OPE coefficient $C_n$.

\subsection{Beyond perturbation theory}
\label{sec:LRI-NP}

Finally, we would like to discuss, very briefly and in parallel with section \ref{sec:SRIconfNP} for the SRI, the prospects of a nonperturbative proof of conformal invariance of the LRI critical point. 

First of all one would have to be convinced that the extra-dimensional defect QFT formulation \reef{eq:S-rew} makes sense beyond perturbation theory.\footnote{For a collection of recent work trying to make nonperturbative sense of various aspects of quantum field theory in fractional number of dimensions see \cite{El-Showk:2013nia,Hogervorst:2014rta,Codello:2014yfa}.} 
 
Alternatively, one can try to get rid of the fractional dimensions altogether by restricting the $y>0$ theory to the radially symmetric sector described by the reduced action \reef{eq:SCSrad}, which is $d+1$ dimensional. This action should still contain the conformal symmetry of the original action. The disadvantage of this formulation is the explicit dependence of the action on $y$. On the other hand it's manifestly nonperturbatively well defined.

Whatever the formulation one uses, one will be reduced to studying the Ward identities for the corresponding stress tensor operator. Analogously to section \ref{sec:SRIconfNP}, one can ask: what if the stress tensor trace contains the following term:
\beq
T^M{}_M\supset \delta^{(p)}(y) \del_\mu v^\mu\ \text{(?)}\,,
\eeq
where $v_\mu$ is a dimension $d-1$ vector operator on the defect---a defect virial current. By assumption, $v^\mu$ is not identically conserved. Like in the discussion following the analogous Eq.~\reef{eq:TK}, we see that unless $v_\mu$ is a total derivative its presence precludes conformal invariance of the correlation functions, while preserving their scale invariance.

So, can our theory contain a defect vector which has dimension exactly $d-1$ and is not a total derivative? Generically, this appears unlikely (with or without the total derivative clause). In perturbation theory, there was no candidate for such an operator (and so it's not surprising that no such term is visible in \reef{eq:TMM1}). To establish this rigorously and nonperturbatively appears as hard as the corresponding problem for the SRI.

We finally note that in the context of \emph{boundary} rather than \emph{defect} quantum field theory the question whether scale invariance implies conformality was recently discussed in \cite{Nakayama:2012ed}.

\section{Conclusions}
The general problem of scale versus conformal invariance is continuing to provide food for thought. In this paper we have discussed this issue for the specific theory corresponding to the LRI at criticality. Since this model most notably does not have a local stress tensor, even the standard (genericity) arguments are invalid, leaving the question of scale versus conformal invariance hanging in midair.

In the first part of the paper we have provided nontrivial evidence for conformal invariance of the LRI by showing that the $\vev{\phi \phi^3}$ and $\vev{\phi^2 \phi^4}$ two point functions vanish at criticality, at least up to terms of $O(\epsilon^3)$. In the second part we were able to \emph{prove} conformal invariance at the critical point to all orders in perturbation theory. The salient point of the proof was the construction of the LRI model as a defect theory in an auxiliary higher-dimensional space; this allowed us to work with a higher-dimensional stress tensor and construct a proof analogous to the one used for the Wilson-Fisher fixed point. Much like the SRI, it is plausible that the LRI is also nonperturbatively conformally invariant.

What does our proof teach us about general long-range lattice models? First of all, the models that have an epsilon expansion starting from a generalized free theory, like the long-range $O(N)$ or Potts models,\footnote{The $q$-component Potts model has a Landau-Ginzburg description via a $(q-1)$-component scalar field with quartic couplings preserving $S_{q}$ symmetry \cite{Zia:1975ha}. This description is trivially generalized to the long-range case. The phase transition is second order if the $S_{q}$-invariant cubic operator is irrelevant at the IR fixed point.} will be conformally invariant at their critical points to all orders by simple extensions of our proof.\footnote{Recall once again that in $1+1$ dimensions conformal invariance without a stress tensor implies a global $SO(3,1)$ invariance, not the full Virasoro symmetry.}

In hindsight it is now also easy to cook up long-range models which will \emph{not} be conformally invariant. The simplest one is based on a vector rather than a scalar. The gaussian long-range action then has two terms consistent with rotation invariance. Equivalently, the vector two point function also has two terms:
\beq
\langle B_{\mu}(x) B_{\nu}(0) = (a \delta_{\mu\nu}+bx^\mu x^\nu/x^2)/{|x|^{2\Delta_B}}\,.
\eeq
This agrees with the two point function of a conformal primary vector for $b/a=-2$, and with that of a derivative of a scalar primary for $b/a=-2\Delta_B$. For all the other ratios this gaussian model will not be conformal \cite{Dorigoni:2009ra}.\footnote{One can also reproduce this theory \emph{\`a la} Caffarelli-Silvestre from an auxiliary vector theory in a higher-dimensional space. The auxiliary theory action has two terms $(\del_M B_N)^2$ and $(\del_M B_M)^2$. As is well known, such a vector theory without gauge invariance (``theory of elasticity") is not conformally invariant \cite{Riva:2005gd,ElShowk:2011gz}.} Let us now perturb the gaussian theory by a quartic local term $(B_\mu B^\mu)^2$ and flow to the IR fixed point. We conjecture that this fixed point will be an interacting scale invariant theory without conformal invariance. It would be interesting to check this by explicit epsilon expansion computations of two point functions, like we did in section \ref{sec:test}. 

In recent years the use of the conformal bootstrap has emerged as an excellent tool to analyze CFTs in various dimensions \cite{Rattazzi:2008pe}. The main ingredients in this approach are the conformal block decomposition of four point functions and the associated crossing symmetry equations, which all follow directly from conformal invariance (see e.g.~\cite{EPFL}). The present work therefore opens the door towards an analysis of the LRI using these methods. Such an analysis would be completely non-perturbative, and is likely to result in a high-precision determination of its critical exponents.

With an eye towards the conformal bootstrap let us finally discuss the critical point of the LRI from the perspective of an ordinary CFT. Firstly, the LRI distinguishes itself from the more familiar, local, CFTs by the absence of a local stress tensor operator. Secondly, we expect a one-dimensional family of solutions to the crossing symmetry, parametrized by the exactly known scaling dimension of $\phi$. 
Thirdly, our analysis uncovered an interesting fact that the second $\bZ_2$ odd primary operator $\phi^3$ is related to $\phi$ through the nonlocal EOM: 
$
\phi^3 \propto \mathcal L_\sigma \phi
$.
This fixes the scaling dimension of $\phi^3$ to be $\Delta_\phi+\sigma$.
Curiously, this can be equivalently rewritten as
\beq
\Delta_{\phi^3} = d - \Delta_\phi\,.
\label{eq:phi13drel-shadow}
\eeq 
This identifies $\phi^3$ as the so-called `shadow' operator of $\phi$, which by definition transforms in a conformal representation with the same value of the quadratic Casimir as $\phi$. The shadow operators often appear in discussions of conformal blocks (see e.g. \cite{Dolan:2000ut}), but usually as a formal tool, since they do not belong to the local operators of the theory. On the contrary, Eq.~\reef{eq:phi13drel-shadow} means that both $\phi$ \emph{and} its shadow are good local operators of the LRI critical point.\footnote{Other examples of theories with pairs of operators satisfying the ``shadow relation'' $\Delta_1+\Delta_2=d$ can be found among SUSY CFTs. Namely, such a relation can emerge if $\calO_1$ is a chiral primary of dimension $d/n$ where $n$ is an integer, and $\calO_2=(\calO_1)^{n-1}$. We thank Leonardo Rastelli for sending us a list of such theories. One simple example is the $\calN=2$ 3d Ising model where $n=3$. What sets LRI apart is that the shadow relation appears as a consequence of the nonlocal EOM.}

One may wonder if the nonlocal EOM also implies an exact proportionality relation between the coefficients with which $\phi$ and $\phi^3$ appear in any OPE. Unfortunately, this appears not to be the case. Using the nonlocal EOM \emph{once}, we can easily relate the three point functions of $\phi^3$ to those of $\phi$. However, we then have to relate the normalizations of the $\phi$ and $\phi^3$ two point functions. This requires using the nonlocal EOM twice, subtracting the unperturbed $\phi$ two point function as in \reef{eq:eomtwice}. The final answer then depends on the relative normalization of $\langle \phi \phi\rangle$ and $\langle\phi \phi\rangle_{S_0}$, which is unknown in general. See appendix \ref{sec:relope}.

We conclude with a brief comment about the limit $\sigma\to\sigma_*$. What happens with the nonlocal CFT describing the LRI critical point when we approach this limit from below? As we discussed in the introductions, for $\sigma>\sigma_*$ the LRI critical point is supposed to be in the SRI universality class. It would be simplest if all correlation functions continuously transitioned to the SRI ones in the limit. However, this seems problematic in view of the presence of $\phi^3$ as a primary in the LRI CFT. 

Consider for example the correlation function $\langle \phi\, \phi^2\, \phi\, \phi^2\rangle$. In the LRI, its conformal block decomposition is expected to contain the contributions of \emph{two} relevant $\bZ_2$ odd scalars: $\phi$ and $\phi^3$. On the other hand, $\varphi^3$ is a descendant in the CFT describing the WF fixed point, while the second $\bZ_2$ odd primary operator ($\varphi^5$) is irrelevant. It could be that $\phi^3$ decouples in the $\sigma\to\sigma_*$ limit, but based on the discussion in appendix \ref{sec:relope} this also seems unlikely. We thus are led to conclude that there is a discontinuity in the transition from the LRI to the SRI correlation functions. It would be very interesting to investigate this in more detail.

\section*{Acknowledgements}

We are grateful to Abdelmalek Abdesselam, Connor Behan, Sheer El-Showk, Pronob Mitter, Hugh Osborn, Jo\~ao Penedones, Leonardo Rastelli, Riccardo Rattazzi and Kay Wiese for the useful discussions and communications. S.R.~is grateful to Sergey Sibiryakov for a discussion which prompted to settle this issue. This research was partly supported by the National Centre of Competence in Research SwissMAP funded by the Swiss National Science Foundation. MFP was supported by a Marie Curie Intra-European Fellowship of the European Community's 7th Framework Programme under contract number PIEF-GA-2013-623606.

\appendix

\section{The hard diagram}
\label{eq:hard}
In this appendix we will discuss the computation of the coefficient $P_{2c}$ in \reef{eq:Pc}, which comes from the last diagram in \reef{eq:phi2phi4diags}. We need to evaluate the integral
\beq
\int {\rm d}^d y\, \d^d z \frac{1}{|x-y|^{2\alpha} }
\frac{1}{|x-z|^{2\alpha} } 
\frac{1}{|y-z|^{2\alpha}}
\frac{1}{|y|^{2\beta}}\frac{1}{|z|^{2\beta}} = \frac{\Upsilon}{|x|^{3d/2-7\eps/2}}\,,
\label{eq:tocomp}
\eeq
where $\alpha=\Delta_\phi=(d-\eps)/4$, $\beta=2\alpha$. We will show that 
   \beq
   \textstyle
\Upsilon=4 \pi ^d {\Gamma \left(-\frac{d}{4}\right)}/{\Gamma
   \left(\frac{3 d}{4}\right)}+O(\eps)\,.
   \label{eq:Ypred}
   \eeq
To get from here to $P_{2c}$ one needs to multiply by the symmetry factor $2^7 3^3/(4!)^2=6$.
    
We will present two methods to compute $\Upsilon$: 
one using Mellin-Barnes representations and another expansions into Gegenbauer polynomials.

\subsection{Mellin-Barnes method}

The $y$ integral in \reef{eq:tocomp} has the following Mellin-Barnes (MB) representation (\cite{Boos:1990rg}, Eq.~(24)):
\bea
I_y&\equiv&\int \ud^d y \frac{1}{|x-y|^{2\alpha}}\,\frac{1}{|y-z|^{2\alpha}}\,\frac{1}{|y|^{2\beta}}\nonumber \\
&=& \frac {\pi^{\frac d2}}{|z-x|^{d-2\epsilon}}\iint \frac{\ud s \ud t}{(2\pi i)^2}\, 
\frac{\mathcal M(s,t)
}
{
\Gamma(\epsilon)\Gamma\left(\frac{d-\epsilon}4\right)^2\,\Gamma\left(\frac{d-\epsilon}2\right)
}
\,\left(\frac{|x|^2}{|z-x|^2}\right)^s\left(\frac{|z|^2}{|z-x|^2}\right)^t\label{ibox}\,,
\label{eq:MB}
\eea
where
\small
\beq
\mathcal M(s,t)=\Gamma(-s)\Gamma(-t)\Gamma\left(-\frac d4+\frac{3\epsilon}4-s\right) \Gamma\left(-\frac d4+\frac{3\epsilon}4-t\right) \Gamma\left(s+t+\frac{d-\epsilon}2\right) \Gamma\left(s+t+\frac{d-2\epsilon}2\right)\,.\quad
\label{eq:Mst}
\eeq
\normalsize
One way to get this representation is to introduce an extra factor $|y-w|^{-2\eps}$ into the integral $I_y$. The resulting integral $\tilde I_y$ is then conformal as the exponents sum to $2d$. Its MB representation is given by Symanzik's formula (Eq.~(5) in \cite{Symanzik:1972wj}). Eq.~\reef{eq:MB} follows by taking the limit $w\to\infty$ in which the conformal cross-ratios in Symanzik's formula tend to $|x|^2/|z-x|^2$ and $|z|^2/|z-x|^2$.

\begin{figure}[htbp]
	\centering
		\includegraphics[width=9cm]{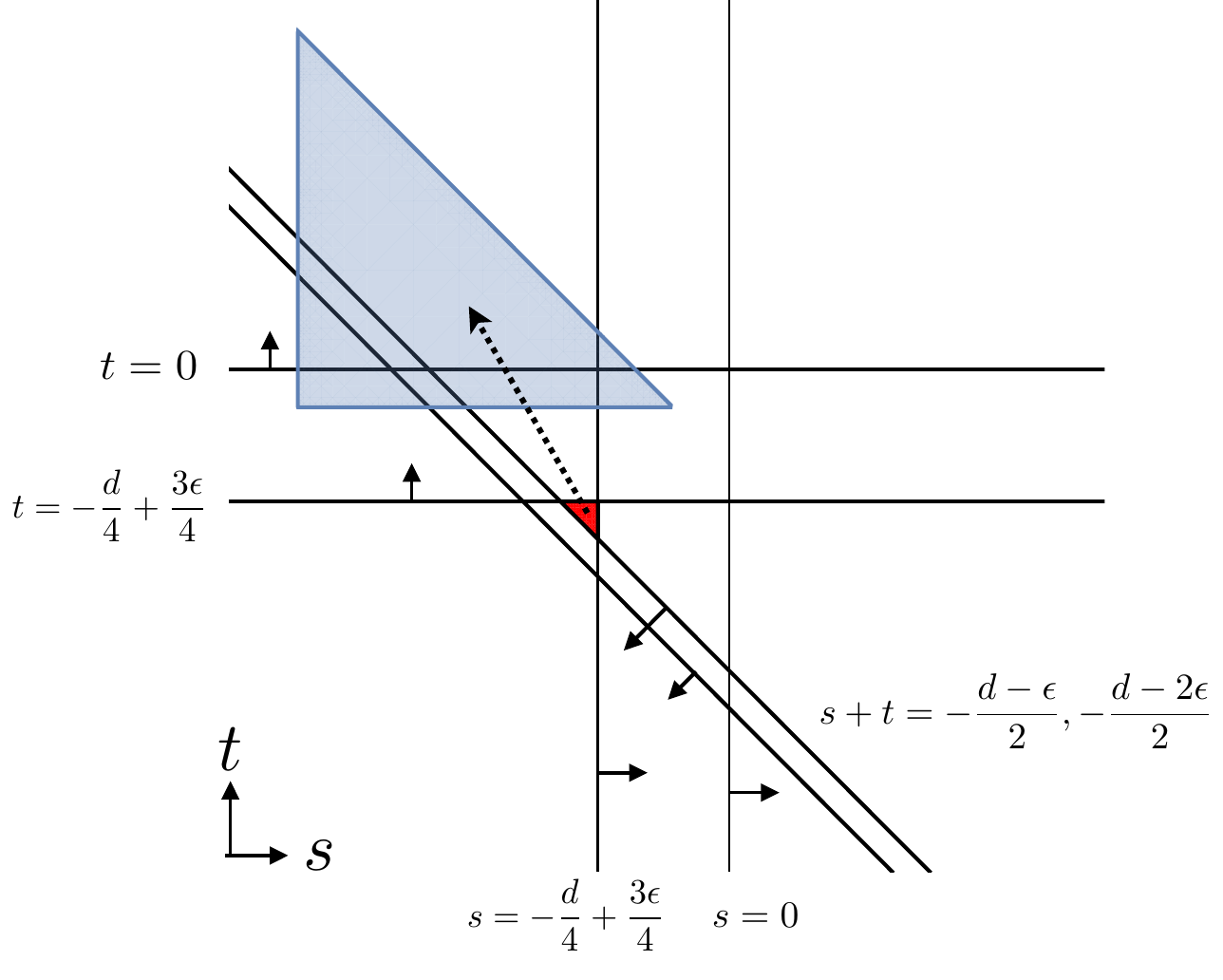}
		\caption{The horizontal, vertical, and diagonal straight lines show the location of the poles in \reef{eq:Mst}. In this plot $d=3$ and $\eps=0.3$, but the same picture is valid for all $d$ of interest and $\eps\ll 1$. Only the leading poles are shown; each line is accompanied by a series of parallel lines spaced by 1 in the direction shown by short arrows. The real parts $(s_0,t_0)$ of the integration contours in the MB representation \reef{eq:MB} must lie in the small red triangle. On the other hand $(s_0,t_0)$ must lie in the large grey triangle for \reef{eq:convord} to be satisfied. The dashed arrow shows a possible contour shift, crossing two horizontal pole lines at $t=-\frac d4+\frac{3\eps}{4}$ and $t=0$.}
	\label{fig-MBpoles}
\end{figure}

The contours in \reef{eq:MB} run from $-i\infty$ to $i\infty$ separating poles of the $\Gamma$ functions whose arguments involve positive and negative $s,t$. We can take straight-line contours $s=s_0+i \bR $ and $t=t_0+i \bR$, where $(s_0,t_0)$ lies in the small red triangle in Fig.~\ref{fig-MBpoles}. E.g.~$s_0=t_0=(-\frac d4+\frac{3\epsilon}4)-\frac\eps8$ will do. That straight line contours can be chosen is related to the fact that the integral $I_y$ is both UV and IR convergent and does not require analytic continuation.

We must now compute the integral over $z$, which is of the form
\bea
\int \ud^d z \frac{1}{|z|^{2\Delta_1}\,|z-x|^{2\Delta_2}}
= \frac{\pi^\frac d2}{\Gamma\left(\Delta_1\right)\Gamma\left(\Delta_2\right)}\, \frac{\Gamma\left(\Delta_1+\Delta_2-\frac d2\right)}{|x|^{2(\Delta_1+\Delta_2-\frac d2)}} \frac{\Gamma(\frac d2-\Delta_1)\Gamma(\frac d2-\Delta_2)}{\Gamma(2h-\Delta_1-\Delta_2)}.
\eea
with
\beq
\Delta_1=({d-\epsilon})/2-t\,,\qquad \Delta_2={3d}/4-{5\epsilon}/4+s+t\,.
\eeq
However, there is an issue. In order to commute the integral over $z$ with the MB integral, we must make sure that the $z$ integral converges in the ordinary sense for all values of $s,t$ in the integration contour. 
The $z$ integral converges for
\bea
0<\mbox{Re}(\Delta_1),\mbox{Re} (\Delta_2)<d/2, \qquad \mbox{Re}(\Delta_1+\Delta_2)>d/2\,,
\label{eq:convord}
\eea
while for the other parameters it's defined by analytic continuation. This condition is violated by the above choice of contour. The solution is to shift the contour before doing the $z$ integral. One possible shift is indicated in Fig.~\ref{fig-MBpoles}. The $t$ contour is shifted up so that it crosses the horizontal line of poles at $t=-\frac d4+\frac{3\eps}{4}$ and $t=0$ and ends up in the bulk of the allowed grey triangular region. The new $t_0$ satisfies the condition $0<t_0<-\frac d4+\frac{3\eps}{4}+1$. 

After this contour shift, we perform the $z$ integration and obtain
\beq
\Upsilon=\frac{\pi^{d}}{\Gamma(\epsilon)\Gamma\left(\frac{d-\epsilon}4\right)^2\,\Gamma\left(\frac{d-\epsilon}2\right)}
(I_1+I_2+I_3)\,,
\label{eq:UMB}
\eeq
where
\begin{align}
I_1&=\iint \frac{\ud s \ud t}{(2\pi i)^2}\, 
\mathcal{ \tilde M}(s,t)\,,
\\
I_2&=-\int \frac{\ud s}{2\pi i}\, 
{\rm Res}_{t=-\frac d4+\frac{3\eps}2}\mathcal{ \tilde M}(s,t)\,,
\\
I_3&=-\int \frac{\ud s}{2\pi i}\, 
{\rm Res}_{t=0}\mathcal{ \tilde M}(s,t)\,,
\\
\mathcal{ \tilde M}(s,t)&=\mathcal M(s,t)\times 
\frac{
\Gamma\left(\frac{3d}4-\frac{7\epsilon}{4}+s\right)\Gamma\left(\frac{\epsilon}2+t\right) \Gamma\left(-\frac d4+\frac{5\epsilon}4-s-t\right)
}
{
\Gamma\left(\frac{d-\epsilon}2-t\right)\Gamma\left(\frac{3d}4-\frac{5\epsilon}4+s+t\right)\Gamma\left(-\frac d4+\frac{7\epsilon}4-s\right)
}\,.
\end{align}
The $I_{2,3}$ come from the pole lines crossed when shifting the contour. Notice that all the $I_i$ integrals converge rapidly as the $\Gamma$'s decay exponentially in the imaginary direction.

Let us now take the $\eps\to0$ limit of \reef{eq:UMB}. Because of the explicit $\Gamma(\epsilon)^{-1}$ prefactor, we are only interested in obtaining the $\epsilon$ poles of the $I_i$'s.

The contour of integration in the double MB integral $I_1$ was chosen serendipitously so that it remains $O(1)$ away from any $\Gamma$ function pole as $\eps\to0$. Because of this $I_1=O(1)$ and does not contribute to the leading asymptotics of $\Upsilon$. 

For the second integral we have:
\beq
I_2\approx \frac{\Gamma \left(\frac{d}{4}\right) 
\Gamma \left(-\frac{d}{4}\right)} {\Gamma \left(\frac{3 d}{4}\right) }
\int \frac{\ud s}{2\pi i}
\frac{[\Gamma (-s)]^2
   \Gamma \left(\frac{3 d}{4}+s\right) 
  }
   {
   \Gamma \left(\frac{d}{2}+s\right) 
   }  
   \times
\frac{
   \Gamma \left(\frac{d}{4}+s-\frac{\epsilon
   }{4}\right) \Gamma \left(\frac{d}{4}+s+\frac{\epsilon
   }{4}\right) \Gamma \left(-\frac{d}{4}-s+\frac{3 \epsilon
   }{4}\right)}
   {
     \Gamma \left(-\frac{d}{4}-s+\frac{7 \epsilon
   }{4}\right)}  \,.
    \eeq
    Here we kept the $\eps$ dependence only in the last group of factors which become singular as $\eps\to0$ because of the position of the contour at $s=-\frac{d}{4}+\frac{3 \epsilon
   }{4}-\frac \eps 8$. By shifting the contour to the right of $s=-\frac{d}{4}+\frac{3 \epsilon
   }{4}$ we can arrange so that the new position of the contour remains $O(1)$ from any pole. Picking up the contribution of the crossed pole, we get:
   \beq
I_2= \frac{\Gamma \left(\frac{d}{4}\right) ^2 \Gamma \left(\frac{d}{2}\right)
\Gamma \left(-\frac{d}{4}\right)} {\Gamma \left(\frac{3 d}{4}\right) } \Gamma(\eps/2)+O(1)\,.
\label{eq:I2}
    \eeq

Apart from an explicit factor of $\Gamma(\eps/2)$, in the rest of $I_3$ we can set $\eps$ to zero. We get:
   \beq
I_3\approx {\textstyle \Gamma\left(\frac\eps 2\right)}\frac{\Gamma \left(-\frac{d}{4}\right)}{\Gamma
   \left(\frac{d}{2}\right)}
 \int { \frac{\ud s}{2\pi i}\textstyle \Gamma (-s) \Gamma
   \left(-\frac{d}{4}-s\right) \Gamma
   \left(\frac{d}{2}+s\right)^2} = {\textstyle \Gamma\left(\frac\eps 2\right)}\frac{\Gamma \left(-\frac{d}{4}\right)}{\Gamma
   \left(\frac{d}{2}\right)}\times \frac{\Gamma \left(\frac{d}{4}\right)^2 \Gamma
   \left(\frac{d}{2}\right)^2}{\Gamma \left(\frac{3
   d}{4}\right)}\,.
   \eeq
    Barnes' lemma was used here to express the integral (the contour in the integral passes between $s=-d/2$ and $s=-d/4$, separating the $\Gamma$ function poles with positive and negative $s$ in the argument). Therefore for $I_3$ we get the same answer as for $I_2$ in \reef{eq:I2}.
    
%
 Finally, plugging the found values of $I_{2,3}$ into \reef{eq:UMB}, we obtain \reef{eq:Ypred}.
 
 \subsection{Gegenbauer polynomial method}

This method expands massless propagators like $1/|x-y|^{2\alpha}$ into powers of $|x|/|y|$ or the inverse, depending on whether $|x|<|y|$ or $|y|<|x|$, times angular factors which are Gegenbauer polynomials. We will use it in a version explained by Kotikov in \cite{Kotikov:1995cw},\cite{Kotikov:2000yd}, where we refer for notation. The key equation is (3) from \cite{Kotikov:2000yd}, which we rewrite as ($\hat x=x/|x|$):
\begin{eqnarray}
\frac{1}{|x-y|^{2\alpha}} ~=~
\frac{1}{(|x||y|)^\alpha} \sum^{\infty}_{n=0} ~C_n^{\alpha}(\hat{x}\hat{y}) T^{\alpha+n},\qquad T=\min(|x|/|y|,|y|/|x|)\,.
  \label{A2a}
\end{eqnarray}
Here the naturally appearing Gegenbauer polynomial is $C_n^{\alpha}$, while it's convenient to have $C_n^{\lambda}$'s with $\lambda=d/2-1$ for subsequent integration. To pass from $\alpha$ to $\lambda$ we
use (5) of \cite{Kotikov:2000yd}:
\begin{eqnarray}
C_n^{\alpha}(t)~=~\sum_{k=0}^{[n/2]}
C_{n-2k}^{\lambda}(t) \frac{(n-2k+\lambda) \Gamma(\lambda)}
{k!\Gamma(\alpha)}  \frac{ \Gamma(n+\alpha-k)\Gamma(k+\alpha-\lambda)}
{\Gamma(n+\lambda+1-k)\Gamma(\alpha-\lambda)}\,.  \label{A7}
\end{eqnarray}
 Substituting back in \reef{A2a} we find, in agreement with (3) in \cite{Terrano:1980af}:
 \beq
\frac{1}{|x-y|^{2\alpha}} =
\frac{1}{(|x||y|)^\alpha}
\frac{ \Gamma(\lambda)}{\Gamma(\alpha)} 
\sum^{\infty}_{n=0}  \frac{\Gamma(n+\alpha)}{\Gamma(n+\lambda)} C_n^{\lambda}(\hat{x}\hat{y}) T^{n + \alpha}
{}_2F_1(n+\alpha,\alpha-\lambda,n+\lambda+1,T^2)\,.
\label{eq:ter}
\eeq
Finally (and this is Kotikov's trick) let us represent Gegenbauers as the products of symmmetric traceless tensors, eq. (7) of \cite{Kotikov:2000yd}:
\beq
C_n^\lambda(\hat x \hat z)=\frac{2^n \Gamma(n+\lambda)}{n!\Gamma(\lambda)} \frac {x^{\mu_1\ldots\mu_n} z^{\mu_1\ldots\mu_n}}
{(|x||z|)^n}\,.
\eeq
We obtain
\begin{align}
\frac{1}{|x-y|^{2\alpha}} &=
\sum^{\infty}_{n=0}  \frac {x^{\mu_1\ldots\mu_n} y^{\mu_1\ldots\mu_n}}
{(|x||y|)^{\alpha+n}} \sum_{k=0}^\infty W^\alpha_{n,k} T^{n+2k + \alpha} \,,
\label{eq:ter1}\\
W^\alpha_{n,k}&= \frac{2^n }{n!}  
\frac{1}{\Gamma(\alpha)} 
 \frac{\Gamma(n+\alpha+k)(\alpha-\lambda)_k}
{k! (n+\lambda+1)_k} \,.
\end{align}

\subsubsection{Integration rules} 

In this section parameters $\alpha$ and $\beta$ are arbitrary, not necessarily the same as in our integral \reef{eq:tocomp}. Denote $D x=\d^dx/(2\pi)^d$, $(n)={\mu_1\ldots\mu_n}$, $\theta_{x,y}=\theta(x^2-y^2)$. We will need the following integrals, see (11),(12),(14),(15) of \cite{Kotikov:2000yd}:
\begin{align}
& \int Dx \frac{x^{(n)}}{x^{2\alpha}(x-y)^{2\beta}} \theta_{x,y}=
\frac{y^{(n)} }{|y|^{2(\alpha+\beta)-d}} M_1(n,\alpha,\beta)\,, \nn \\
& \int Dx \frac{x^{(n)}}{x^{2\alpha}(x-y)^{2\beta}} \theta_{y,x}=
\frac{y^{(n)} }{|y|^{2(\alpha+\beta)-d}} M_2(n,\alpha,\beta) \,,
\label{eq:ints}
\end{align}
and
\begin{align}
& \int Dx \frac{x^{(n)}}{x^{2\alpha}(x-y)^{2\beta}} \theta_{x,z}=y^{(n)} 
\Biggl\{\frac{\theta_{y,z}}{|y|^{2(\alpha+\beta)-d}} K(n,\alpha,\beta)\nn\\
&+\sum_{m=0}^\infty \frac{1}{|z|^{2(\alpha+\beta)-d}}
\Biggl[
\left(\frac{y^2}{z^2}\right)^m L_1(m,n,\alpha,\beta)\, \theta_{z,y}
-\left(\frac{z^2}{y^2}\right)^{m+\beta+n} L_2(m,n,\alpha,\beta)\,\theta_{y,z}
\Biggr ]
\Biggr\}\,,\nn\\
&\int Dx \frac{x^{(n)}}{x^{2\alpha}(x-y)^{2\beta}} \theta_{z,x}=y^{(n)} 
\Biggl\{\frac{\theta_{z,y}}{|y|^{2(\alpha+\beta)-d}} K(n,\alpha,\beta)\nn\\
&-\sum_{m=0}^\infty \frac{1}{|z|^{2(\alpha+\beta)-d}}
\Biggl[
\left(\frac{y^2}{z^2}\right)^m L_1(m,n,\alpha,\beta) \theta_{z,y}
-\left(\frac{z^2}{y^2}\right)^{m+\beta+n} L_2(m,n,\alpha,\beta)\theta_{y,z}
\Biggr]
\Biggr\}\,.
\label{eq:intc}
\end{align}
Here ($\omega=1/{(4\pi)^{d/2} }$):
\begin{gather}
K(n,\alpha,\beta)=\omega A^{n,0}(\alpha,\beta)\,,\nn\\
L_1(m,n,\alpha,\beta)=\omega  \frac{B(m,n|\beta,\lambda)}{m+\alpha+\beta-d/2}\,,\quad
M_1(n,\alpha,\beta)=\sum_{m=0}^\infty L_1(m,n,\alpha,\beta)\,,\\
 L_2(m,n,\alpha,\beta)=\omega  \frac{B(m,n|\beta,\lambda)}{m-\alpha+n+d/2}\,,\quad
M_2(n,\alpha,\beta)=\sum_{m=0}^\infty L_2(m,n,\alpha,\beta)\,,\nn
\end{gather}
while $A^{n,0}(\alpha,\beta)$ and $B(m,n|\beta,\lambda)$ are given by
\begin{gather}
A^{n,0}(\alpha,\beta)=\frac{a_n(\alpha) a_0(\beta)}{a_n(\alpha +\beta -d/2)},\qquad a_n(\alpha)=\frac{\Gamma(d/2-\alpha+n)}{\Gamma(\alpha)}\,,\nn\\
B(m,n,\beta,\lambda)=\frac{\Gamma(m+n+\beta) (\beta-\lambda)_m}{m!\Gamma(m+n+d/2)}\,.
\end{gather}
It's possible to perform the summation in the definitions of $M_{1,2}$, with the result expressible in terms of ${}_3F_2$'s at $z=1$:
\begin{align}
M_1(n,\alpha,\beta)=&-\frac{2^{1-d} \pi ^{-d/2} \Gamma (n+\beta ) }{\Gamma (\beta ) (-2 \alpha -2 \beta +d) \Gamma
   \left(\frac{1}{2} (d+2 n)\right)}\nn\\
   &\times  _3F_2\left(-d/2+\beta +1,n+\beta ,-d/2+\alpha
   +\beta ;d/2+n,-d/2+\alpha +\beta
   +1;1\right)\,,\nn\\ 
M_2(n,\alpha,\beta)= & \frac{2^{1-d} \pi ^{-d/2} \Gamma (n+\beta ) }{\Gamma (\beta )
   (-2 \alpha +d+2 n) \Gamma \left(\frac{1}{2} (d+2 n)\right)}\\
   &\times \,
    _3F_2\left(d/2+n-\alpha ,-d/2+\beta +1,n+\beta
   ;d/2+n,d/2+n-\alpha +1;1\right)\,.\nn
   \end{align}
   
\subsubsection{Computation} 
 Let's start with the integral in $y$:
 \beq
\int Dy \frac{1}{|x-y|^{2\alpha}|y-z|^{2\alpha}|y|^{2\beta}} \,.
\eeq
We will expand the $|x-y|$ propagator. (We checked that expanding the $|y-z|$ propagator gives the same result.) After the expansion we get
\beq
\int Dy  \sum_{n,k} x^{(n)} y^{(n)} W_{n, k}^\alpha |y-z|^{-2\alpha} (|x|^{-2 \alpha - 2 k - 2 n} |y|^{-4 \alpha + 2 k}
    \theta_{x,y}  + |x|^{2 k} |y|^{-6 \alpha - 2 k - 2 n}  \theta_{x,y})\,.
    \label{eq:int1}
\eeq
All the integrals can be done using Eq.~\reef{eq:intc}, producing terms of the form $x^{(n)} z^{(n)}$
times $\theta_{x,z}$ or $\theta_{z,x}$ times powers of $|x|$ and $|z|$. We will then multiply this by
\beq
 {|x-z|^{-2\alpha} |z|^{-2\beta}}
 \label{eq:multiply}
\eeq
and do $\int Dz$, using \reef{eq:ints}. The arising products $x^{(n)} x^{(n)}$ can be reduced using (7) of \cite{Kotikov:2000yd}:
\beq
x^{(n)} x^{(n)} = f_n |x|^{2n},\quad f_n= 
{(2\lambda)_n}/[{2^n (\lambda)_n}]\,.
\eeq
The end result will collapse to an infinite series of terms all proportional to $|x|$ to the same power, and we have to sum the series to get the prefactor.

Let's supply the details. Integrating the $\theta_{x,y}$ term in \reef{eq:int1} we get 
\begin{align}
 \sum_{n,k} &x^{(n)} z^{(n)} W_{n, k}^\alpha \biggl\{|x|^{-2 \alpha - 2 k - 2 n} |z|^{-6 \alpha + d + 2 k}
   K(n, 2 \alpha - k, \alpha)  \theta_{x, z}\nn\\ &- 
\sum_m |x|^{-8 \alpha + d - 2 m - 2 n} |z|^{2 m}
   L_1(m, n, 2 \alpha - k, \alpha)  \theta_{x, z}\nn\\&+ 
\sum_m |x|^{-6 \alpha + d + 2 m} |z|^{-2 \alpha - 2 m - 2 n}
   L_2(m, n, 2 \alpha - k, \alpha)  \theta_{z, x}\biggr\}
   \label{eq:int2}\,.
\end{align}
Now let's specialize to $\alpha=(d-\eps)/4$. We will only be interested in extracting the leading dependence in the $\eps\to0$ limit. As we will see, the end results is finite in this limit: $1/\eps$ poles which appear at the intermediate steps of the computation cancel or get multiplied by $O(\eps)$ coefficients at later stages. 

It's interesting to trace how these potential $1/\eps$ terms disappear. From \reef{eq:int1}, we can expect a $1/\eps$ dependence in the integral of the $\theta_{x,y}$ term when $k=n=0$. In \reef{eq:int2}, this divergence is present in the $K$ terms for $n=k=0$ and in the $L_2$ term for $n=k=m=0$, with the same leading coefficient. Gathering these divergent terms, we have:
\beq
\frac{2^{1 - d} \pi^{-d/2}}{\Gamma(d/2)\eps} |x|^{-2 a}|z|^{-2a}(|z|^\eps \theta_{x,z} + |x|^\eps \theta_{z,x})\,.
\eeq
This grouping makes it easy to see that in the subsequent integration in $z$ these terms will be multiplied by an $O(\eps)$ coefficient, giving an $O(1)$ contribution to the final answer. This can be checked explicitly in the formulas below.

Analogously, the integral of the $\theta_{y,x}$ terms in \reef{eq:int1} gives:
\begin{align}
\sum_{n,k} & x^{(n)} z^{(n)} W_{n, k}^a \biggl\{
|x|^{2 k} |z|^{-8 \alpha + d - 2 k - 2 n}
   K(n, 3 \alpha + k + n, \alpha) \theta_{z, x}\nn\\
   &+\sum_m
   |x|^{-8 \alpha + d - 2 m - 2 n} |z|^{2 m}
   L_1(m, n, 3 \alpha + k + n, \alpha)\theta_{x, z}\nn\\  
    &- \sum_m
 |x|^{-6 \alpha + d + 2 m} |z|^{-2 \alpha - 2 m - 2 n}
   L_2(m, n, 3 \alpha + k + n, \alpha)\theta_{z, x}\biggr\}\,.
   \label{eq:int3}
   \end{align}
 None of these terms are singular in the $\eps\to0$ limit.
 
Summing \reef{eq:int2} and \reef{eq:int3}, multiplying by \reef{eq:multiply} and doing $\int Dz$, we obtain:\footnote{The factor $(2\pi)^{2d}$ comes from the definition of $Dx$.}
\begin{align}
\Upsilon=(2\pi)^{2d}
   \sum_{n,k} f_n W_{n, k}^\alpha
\Bigl\{ & [K(n, 2 \alpha - k, \alpha) M_2(n, 5 \alpha - d/2 - k, \alpha)
\nn\\[-7pt]
&+ K(n, 3 \alpha + k + n, \alpha) M_1(n, 6 \alpha - d/2 + k + n, \alpha)]\nn\\
+ \sum_{m}&[ 
   L_2(m, n, 2 \alpha - k, \alpha) M_1(n, 3 \alpha + m + n, \alpha)\nn\\[-7pt]
      & - 
   L_2(m, n, 3 \alpha + k + n, \alpha) M_1(n, 3 \alpha + m + n, \alpha)\nn \\
   &- L_1(m, n, 2 \alpha - k, \alpha) M_2(n, 2 \alpha - m, \alpha)\nn\\
  & + 
   L_1(m, n, 3 \alpha + k + n, \alpha) M_2(n, 2 \alpha - m, \alpha)]\Bigr\}\,.
 \label{int4}  \end{align}
   Among the $M_1$ and $M_2$ coefficients here, the only one which can become singular as $\eps\to0$ is $M_2(n,2a-m,\alpha)$. The terms with this coefficient came from the integration of the second terms in \reef{eq:int2} and \reef{eq:int3}. The singularity happens only for $m=n=0$, and is given by:
\beq
-\frac{2^{-2 d} \pi ^{-d} \Gamma
   \left(-\frac{d}{4}+k+1\right) \Gamma
   \left(\frac{d}{4}+k\right)}{\epsilon  k! \Gamma
   \left(1-\frac{d}{4}\right) \Gamma \left(\frac{d}{4}\right) \left[\Gamma
   \left(\frac{d}{2}\right)\right]^2
   \left(\frac{d}{2}\right)_k}
   \left[-\frac{2}{d-4 k}+ \frac{1}{d+2 k}\right]\,.
\eeq
{\sc Mathematica} sums the series in $k$ to a product of $\Gamma$ functions, {\tt FullSimplify}'ed to zero.

Denote the terms associated with the six lines of \reef{int4} as $C^{(i)}$, $i=1\ldots6$. We have:
\begin{align}
\lim_{\eps\to0}&\Upsilon =I_1+I_2+J_1+J_2+J_3+J_4\,,
\label{eq:6terms}
\end{align}
\begin{align}
&I_1= C^{(1)}_{k=n=0}+C^{(3)}_{k=n=m=0}|_{\eps\to0}\,,\nn\\
&I_2= \sum\nolimits_{k} [C^{(5)} +C^{(6)}]_{n=m=0}|_{\eps\to0}\,,\nn\\
&J_1=\sum\nolimits_{(k,n)\ne (0,0)} C^{(1)} |_{\eps\to0} \,,\nn\\
&J_2=\sum\nolimits_{(k,n,m)\ne (0,0,0)} C^{(3)}|_{\eps\to0} \,,\nn\\ 
&J_3=\sum\nolimits_{k,n,m} C^{(4)}|_{\eps\to0}\,, \nn\\
&J_4=\sum\nolimits_{k,n,m:(n,m)\ne (0,0)} C^{(5)} +C^{(6)}|_{\eps\to0}\,.\nn
\end{align}
We omitted $C^{(2)}$ as it vanishes for $\eps\to0$. We also took into account the above discussion about the separation of the $1/\eps$ poles. The $I_1$ is finite for $\eps\to0$. In $I_2$, the individual terms are $\sim 1/\eps$ with coefficients which sum to zero as we mentioned above. So we can drop these singular parts and sum the $O(\eps^0)$ pieces. The $J_i$ series are manifestly finite as $\eps\to0$. 

We used \reef{eq:6terms} to perform a numerical check of \reef{eq:Ypred} for a few values of $d$.
Although the representaion does not converge very rapidly, including 200 terms in the single series and terms with $n,k,m$ up to 40 in the double and triple series, we found an agreement within one percent or better. This check complements the MB result from the previous section.
%
%
%
%
%
%
%
%
%
%
%
%
%

\section{Conformal invariance of composite operators}
\label{sec:ope}

We would like to discuss to what extent conformal invariance of the correlators of $\varphi$ can be used to show conformal invariance of the composite operators, as mentioned in section \ref{sec:SRIconf}.

We will use the fact that composite operators can be obtained by fusing $\varphi$'s via the OPE. The OPE is well established in perturbation theory \cite{Collins:1984xc}. Here we will use it in the IR where all operators acquire their fixed point dimensions. So we really set $\lambda=\lambda_*$. Let's start by considering a simple example---the operator $\varphi^2$.\footnote{In this section all operators will be finite, renormalized, so we drop square brackets from the notation.} In the IR region the relevant OPE takes the form ($\Delta\equiv\Delta_\varphi$, $\Delta_2\equiv\Delta_{\varphi^2}$):\footnote{The value of the constant coefficient with which $\varphi^2$ appears in this OPE is unimportant for this argument, so we set it to one.}
\beq
\varphi(x) \varphi(0)=|x|^{-2\Delta}(1+ |x|^{\Delta_2}\varphi^2(0)+\ldots)\,.
\label{eq:phixphi}
\eeq
This equation can be used to determine correlation functions of $\varphi^2$ if correlators of $\varphi$ are known.
Indeed consider the correlator
\beq
\langle \varphi(x_1)\varphi(x_2)\ldots\rangle\,.
\eeq
We then have:
\beq
\langle \varphi^2(x_1)\ldots\rangle=\lim_{x_{2}\to x_1} |x_{12}|^{2\Delta-\Delta_2}(\langle \varphi(x_1)\varphi(x_2)\ldots\rangle-|x_{12}|^{-2\Delta}\langle \ldots\rangle)\,.
\label{eq:phi2def}
\eeq
This limit should be understood as $x_{1,2}$ becoming much closer to each other than to any other insertion.

Now, Eq.~\reef{eq:phi2def} implies that correlators of $\varphi^2$ are conformally covariant. Indeed, under a conformal transformation $x\to f(x)$ both terms in the RHS acquire the same factor
\beq
[J(x_1)J(x_2)]^{-\Delta_2/2}\approx J(x_1)^{-\Delta_2}\,,
\eeq
where $J=|\del f/\del x|^{1/D}$ is the $x$-dependent scale factor. This is precisely the transformation rule of a primary of dimension $\Delta_2$.

The above argument can be repeated by fusing other pairs of $\varphi$'s. This proves that correlation functions of an arbitrary number of $\varphi$'s and an arbitrary number of $\varphi^2$'s are conformally covariant. 

The next step is to discuss the derivatives of $\varphi^2$, which appear among the $\ldots$ terms in \reef{eq:phixphi}.
We claim that the coefficients of these terms are such that this part of the OPE respects conformal invariance.
To prove this consider the three point function
\beq
\langle \varphi^2(y) \varphi(x_1)\varphi(x_2)\rangle\,.
\eeq
By the proven conformal invariance property, it has the well-known functional form \cite{Polyakov:1970xd}
\beq
\frac{const.}{|x_1-y|^{\Delta_{2}}|x_2-y|^{\Delta_{2}}|x_1-x_2|^{2\Delta-\Delta_2}}\,.
\eeq
Expanding this in $x_2-x_1$, we will get a series of the schematic form
\beq
\frac{const.}{|x_1-y|^{2\Delta_{2}}}|x_1-x_2|^{2\Delta-\Delta_2}\sum_{n=0}^\infty c_n \left(\frac{x_2-x_1}{|x_1-y|}\right)^n\,.
\label{eq:serexp}
\eeq
As is well known, matching this series to the one obtained from the OPE, we can fix all the coefficients of the $\del^\alpha \varphi^2$ terms, consistently with conformal invariance. Moreover this proves that $\varphi^2$ has zero two point functions with any other operator $\calO$ appearing in the OPE $\varphi\times\varphi$. Indeed, a nonzero two point function would give rise to a term scaling as $1/|x_1-y|^{\Delta_{2}+\Delta_\calO}$. Assuming as we will, that in a generic interacting theory such as the WF fixed point all nontrivial operator degeneracies are lifted, there is no room for such a term in \reef{eq:serexp} except if $\calO$ is a derivative of $\varphi^2$.

Let us then consider the next operator in the $\varphi\times\varphi$ OPE which is not among the already considered operators $\unit$, $\varphi^2$, and the derivatives of $\varphi^2$. It will be the operator $\varphi^4$. Its correlation functions can be defined from those of $\varphi$ by the equation:
\begin{align}
\langle \varphi^4(x_1)\ldots\rangle=\lim_{x_{2}\to x_1} |x_{12}|^{2\Delta-\Delta_4}&\Bigl(\langle \varphi(x_1)\varphi(x_2)\ldots\rangle
\nn\\
&-|x_{12}|^{-2\Delta}\langle \ldots\rangle
\nn\\
&-|x_{12}|^{-2\Delta+\Delta_2}\langle P(x_2-x_1,\del)\varphi^2(x_1)\ldots\rangle\Bigr)\,.
\label{eq:phi4def}
\end{align}
Here $P$ is the infinite series encapsulating the contributions of $\varphi^2$ and of its descendants in the OPE.\footnote{Only the first few terms of this series, corresponding to the descendants of dimension $<\Delta_4$, will contribute in this equation.}
This part of the OPE has already been proven to be conformally invariant. It is then not difficult to convince oneself that all three terms in the RHS of \reef{eq:phi4def} transform under conformal transformations as expected for a primary of dimension $\Delta_4$.

It should be clear that the above argument can be continued recursively to higher dimensions. At each step we will identify a new operator, show that it's a primary, and then show that contributions of all of its descendants in the OPE are consistent with conformal invariance. We then subtract away this part of the OPE, identify the next operator, and go on. In the above examples the considered primaries $\varphi^2$ and $\varphi^4$ were scalars, but modifications to tensor operators are straightforward. 

The above argument thus determines conformal transformation properties of all operators in the $\varphi\times\varphi$ OPE. It is expected that all $\bZ_2$ even primary operators which are scalars or symmetric traceless tensors of even spin appear in this OPE. $\bZ_2$ odd operators can then be accessed via the OPE $\varphi\times \varphi^2$. Mixed Lorentz symmetry representations can be accessed by considering OPEs of operators with spin. This will be somewhat more complicated since there are several conformally covariant structures for the three point functions of operators with spin. This complication is only technical, and the above argument can be suitably extended. We will not discuss this in detail.

\section{Relative normalization of the \texorpdfstring{$\phi$}{phi} and \texorpdfstring{$\phi^3$}{phi3} OPE coefficients}
\label{sec:relope}
In this appendix we investigate the consequence of the nonlocal equation of motion \eqref{eq:nleom} for OPE coefficients and operator normalizations. We will be working in the IR theory. With a small redefinition of the constants involved we can write \eqref{eq:nleom} as
\be
\label{eq:nleomapp}
\int d^d y \frac{1}{|x-y|^{2(d - \D_\phi)}} \phi(y) = C\phi^3(x)\,,\qquad C=-\frac{g_0}{3!} (2\pi)^{2d}w_{d-\sigma}w_{d+\sigma}>0\,. 
\ee

Let us first investigate the consequences of the equation of motion for three point functions. Using \eqref{eq:nleomapp}, and Symanzik's star integral formula \cite{Symanzik:1972wj}, we can easily deduce that if
\be
\vev{\phi(x) \op_1(y) \op_2(z)} = \frac{\lambda_{12\phi}}{|x-y|^{\D_\phi + \D_1 - \D_2} |x-z|^{\D_\phi - \D_1 + \D_2} |y-z|^{- \D_\phi + \D_1 + \D_2}}\,,
\ee
then
\be
\vev{\phi^3(x) \op_1(y) \op_2(z)} = \frac{\lambda_{12 \phi^3}}{|x-y|^{\D_{\phi^3} + \D_1 - \D_2} |x-z|^{\D_{\phi^3} - \D_1 + \D_2} |y-z|^{- \D_{\phi^3} + \D_1 + \D_2}}\,,
\ee
with $\D_{\phi^3} \equiv d - \Delta_\phi$ and with a relative three point function coefficient given by
\be
\frac{\lambda_{12\phi^3}}{\lambda_{12 \phi}}= \frac{M_3}{C}\,,\qquad
M_3 = \pi^{d/2}\frac{\G(\hf(d - \Delta_\phi + \Delta_{12})) \G(\hf(d - \Delta_\phi - \Delta_{12})) \G(\Delta_\phi - d/2)}{\G(d-\Delta_\phi)\G(\hf(\Delta_\phi + \Delta_{12})) \G(\hf(\Delta_\phi - \Delta_{12}))}\,.
\ee
Here we introduced $\Delta_{12} \equiv \Delta_1 - \Delta_2$. Notice that this computation provides an alternative derivation of the scaling dimension $\Delta_{\phi^3}$ in the IR theory, which was derived from the two point function in the main text. Notice also that the gaussian limit $\epsilon \to 0$ is smooth because in this limit $\lambda$ is only nonzero if $\D_\phi = \pm \Delta_{12} - 2k$, and the zero in $M_3$ precisely cancels the overall $1/g_0\sim 1/\epsilon$.

The above three point functions are rather meaningless without knowledge of the two point functions of the operators involved. The two point function of $\phi$ takes the form
\be
\vev{\phi(x) \phi(0)} = \frac{1 + \epsilon^2 \rho}{|x|^{2\D_\phi}}\,.
\ee
The coefficient $\rho$ is a nontrivial function of $\epsilon$ which is finite as $\epsilon \to 0$. The two point function of $\phi^3$ is then given by the analogue of \eqref{eq:eomtwice}, which in the current conventions takes the form
\be
\vev{\phi^3(x) \phi^3(0)} = \int d^d y \int d^d z\, \frac{\eps^2\rho / C^2}{|x-y|^{2(d-\D_\phi)}|z|^{2 (d - \D_\phi)}|y-z|^{2 \D_\phi}}\,.
\ee
Notice that we subtracted the tree-level part from the two point function of $\phi$, cf.~the discussion below equation \eqref{eq:eomtwice}. The resulting two point function is therefore proportional to $\rho$. The integral as written is actually singular, but we can view it as a formal expression that has a well-defined and finite meaning in momentum space. It then evaluates to
\be
\label{eq:phi3phi3app}
\vev{\phi^3(x) \phi^3(0)} = \frac{\eps^2 \rho M_2 / C^2}{|x|^{2 \D_{\phi^3}}} 
\ee
with
\be
\label{eq:m2}
M_2 = \frac{\pi^d \G(d/2 - \D_\phi) \G(\D_\phi - d/2)}{\G(d-\D_\phi)\G(\D_\phi)}\,.
\ee
We emphasize that the gaussian limit is again smooth.

Substituting $\D_\phi = (d- \epsilon)/4$ in \eqref{eq:m2} we find that $M_2 < 0$ in a finite interval around $\epsilon = 0$, for all physical values of $d$. The coefficient $\rho$ is then also expected to be negative, so that altogether the coefficient of \eqref{eq:phi3phi3app} comes out positive. Indeed we expect the LRI to be described by a unitary, or reflection positive in the Euclidean, conformal field theory.\footnote{The gaussian theory is unitary as long as $\Delta_\phi$ is above the scalar unitarity bound $d/2-1$, i.e. if $\sigma<2$. This condition is satisfied throughout region 2, see Fig.~\ref{fig-standard}. We expect that perturbing a unitary theory by a hermitian operator $\phi^4$ with a real coupling will give rise to a unitary theory.} We see from \reef{eq:Q} that $\rho$ is negative to the first nontrivial order in perturbation theory, which is in agreement with expectations.

Meaningful three point functions involve the unit normalized operators
\be
\tilde \phi(x) \equiv \frac{1}{\sqrt{1 + \epsilon^2 \rho}} \phi(x)\,, \qquad \qquad \tilde \phi^3(x) \equiv \frac{1}{\sqrt{\eps^2\rho M_2 / C^2}} \phi^3(x)\,,
\ee
in terms of which we find that
\be
\label{eq:relative3pt}
\frac{\lambda_{12 \tilde \phi^3}}{\lambda_{12 \tilde \phi}} = \frac{M_3 \sqrt{1 + \epsilon^2 \rho}}{\sqrt{\epsilon^2 \rho M_2}}\,.
\ee
We see that $C$ drops out, but the relative three point functions still involve $\rho$ which we can only compute perturbatively. 

Notice however that we can take further ratios to get rid of the unknown $\rho$. Namely, if we consider two OPEs $\calO_1\times\calO_2$ and $\calO_1'\times\calO_2'$, then 
\be
\label{eq:relative3pt1}
\frac{\lambda_{12 \tilde \phi^3}/\lambda_{12 \tilde \phi}}
{\lambda_{1'2' \tilde \phi^3}/\lambda_{1'2' \tilde \phi}} = \frac{M_3}{M_3'}\,.
\ee
This equation may be used in a variety of ways in conformal bootstrap analyses involving multiple correlators.

It is interesting to consider the limit $\sigma \to \sigma_*$ where the LRI should transition to the SRI. The SRI does not have an operator corresponding to our $\tilde \phi^3$. However from \eqref{eq:relative3pt} we see that there is no decoupling of the $\tilde \phi^3$ operator unless $\epsilon^2 \rho = -1$, which would imply that $\phi$ becomes null and also decouples. Barring this decoupling, we predict a discontinuity in the transition from LRI to 
SRI.\footnote{{\bf Note added:} Our findings seem consistent with the scenario proposed in \cite{Parisi}, where it is argued that the LRI/SRI transition is continuous, and that the $\phi$ two point function normalization in the IR vanishes at the transition point. We thank Giorgio Parisi for pointing out this paper.}
\small
\bibliography{LRI-biblio.bib}
\bibliographystyle{utphys.bst}
\end{document}